%% file: main.tex
\documentclass[acmsmall,screen]{acmart}

\AtBeginDocument{%
  }

\setcopyright{acmlicensed}
\copyrightyear{2018}
\acmYear{2018}
\acmDOI{XXXXXXX.XXXXXXX}
\acmConference[Conference acronym 'XX]{Make sure to enter the correct
  conference title from your rights confirmation email}{June 03--05,
  2018}{Woodstock, NY}
\acmISBN{978-1-4503-XXXX-X/2018/06}

\input{macro}

\begin{document}

\title{Evaluating Code Reasoning Abilities of Large Language Models Under Real-World Settings}

\author[Changshu Liu, Alireza Ghazanfari, Yang Chen, Reyhaneh Jabbarvand]{Changshu Liu, Alireza Ghazanfari, Yang Chen, Reyhaneh Jabbarvand}

\begin{abstract}

Code reasoning tasks are becoming prevalent in large language model (LLM) assessments. Yet, there is a dearth of studies on the impact of real-world complexities on code reasoning, e.g., inter- or intra-procedural dependencies, API calls, deeply nested constructs, and non-primitive complex types. Evaluating LLMs under such a simplistic setting poses a significant threat to assumptions about their generalizability in practice. To enable a more realistic evaluation of code reasoning, we collect a dataset of \num{1200} reasoning problems, \name, from two sources: existing code reasoning benchmarks and popular GitHub Python repositories. Our pipeline 
leverages static and dynamic program analysis to automatically serialize/deserialize compound, complex, and custom types galore in real-world code, going far beyond only primitive types used in prior studies. A key feature of \name is categorizing each reasoning problem as \emph{Lower Complexity (LC)} or \emph{Higher Complexity (HC)} via a principled majority-vote mechanism over \emph{nine} diverse and interpretable code-complexity metrics, yielding two well-separated, semantically meaningful categories of problem difficulty suitable for precise calibration of LLM reasoning ability. This categorization shows that the problems used in existing code-reasoning evaluation mostly belong to the \emph{LC} category, failing to represent real-world complexity.

A comprehensive evaluation of \emph{ten LLMs} (including reasoning and general LLMs) on four code reasoning tasks---input, output, loop, and branch prediction---using \name-lite ($500$ reasoning problems) shows a significant performance drop from \emph{LC} to \emph{HC} problems ($37.36\%$, $36.16\%$, $20.90\%$, and $48.60\%$ for input, output, loop, and branch prediction), confirming that conclusions drawn from prior studies using \emph{LC} problems are inflated or unrealistic concerning the reasoning capabilities of LLMs. 
Our pipeline integrates static analysis algorithms and visualization tools to assess the impact of various programming constructs, call chain length, prompting/training strategies, overall program complexity, and individual complexity metrics on LLMs' reasoning abilities. These insights, along with a systematic approach to reasoning failure analysis, result in a taxonomy of $18$ reasoning failure categories that can help develop future LLMs for better code reasoning. 
\end{abstract}

\maketitle

\input{Sections/Introduction}
\input{Sections/Method}
\input{Sections/Experiment}
\input{Sections/Related-Work}

\input{Sections/Conclusion}


\balance
\bibliographystyle{ACM-Reference-Format}
\bibliography{reference}

\end{document}

%% file: macro.tex
\usepackage[utf8]{inputenc} 
\usepackage[T1]{fontenc}    
\usepackage{hyperref}       
\usepackage{url}            
\usepackage{booktabs}       
\usepackage{amsfonts}       
\usepackage{nicefrac}       
\usepackage{microtype}      
\usepackage{xcolor}         
\usepackage{algpseudocode}
\usepackage{lipsum}
\usepackage{xspace}
\usepackage{fancyvrb}
\usepackage{graphicx}
\usepackage{caption}
\usepackage{soul}
\usepackage{adjustbox}
\usepackage{multirow}
\usepackage{float}
\usepackage{comment}
\usepackage[dvipsnames]{xcolor}
\usepackage{wrapfig}
\usepackage{enumitem}
\usepackage{tabularx}
\usepackage{float}
\usepackage{amsmath}
\usepackage{cleveref}
\usepackage{stmaryrd} 
\usepackage{balance} 
\usepackage[table,xcdraw,dvipsnames]{xcolor}
\usepackage[many]{tcolorbox}
\usepackage{makecell}
\usepackage{bm}
\usepackage{algorithm}
\usepackage{algpseudocode}
\usepackage{siunitx}
\usepackage{booktabs,makecell}
\usepackage{pifont}
\algrenewcommand\algorithmicrequire{\textbf{Input:}}
\algrenewcommand\algorithmicensure{\textbf{Output:}}
\algrenewcommand{\algorithmiccomment}[1]{\hspace{.5em}\(\triangleright\)\,#1}
\algtext*{EndIf}
\algtext*{EndFor}
\algtext*{EndWhile}

\newcolumntype{R}[2]{%
    >{\adjustbox{angle=#1,lap=\width-(#2)}\bgroup}%
    l%
    <{\egroup}%
}

\newcommand{\SweB}{SWE-bench\xspace}
\newcommand{\real}{real-world projects\xspace}
\newcommand{\Real}{Real-world Projects\xspace}
\newcommand{\crux}{CRUXEval\xspace}
\newcommand{\heval}{HumanEval\xspace}
\newcommand{\avatar}{Avatar\xspace}
\newcommand{\ceval}{ClassEval\xspace}
\newcommand{\codemind}{CodeMind\xspace}
\newcommand{\reval}{REval\xspace}
\newcommand{\coco}{CoCoNUT\xspace}
\newcommand{\name}{\textsc{RE2-Bench}\xspace}

\newcommand{\codellama}{CodeLlama\xspace}

\newcommand{\dsv}{DeepSeek-V3.2\xspace}
\newcommand{\gfmini}{GPT-5-mini\xspace}
\newcommand{\geminip}{Gemini-3-Pro\xspace}
\newcommand{\haiku}{Claude-Haiku-4.5\xspace}
\newcommand{\cwm}{CWM\xspace}

\newcommand{\highComp}{\textsc{H}\xspace}

\newcommand{\lowComp}{\textsc{L}\xspace}

\newcommand{\changshu}[1]{\textcolor{ForestGreen}{\textbf{Changshu:} #1}}

\definecolor{myhighlight}{HTML}{FCFF2F}

%% file: Sections/Introduction.tex
\section{Introduction}
\label{sec:introduction}

Large Language Models (LLMs) have shown promise in various programming tasks~\cite{ding2024cycle,liu2023your, li2025s,yang2024swe,jimenez2023swe,yuan2024evaluating,ryan2024code,pan2024lost,ibrahimzada2024repository,Ding2025Vulnerability,shestov2025finetuning}, including code reasoning~\cite{liu2025assessing,liu2024codemind,gu2024cruxeval, liu2025tool, chen2024runtime,beger2025coconut}. Code reasoning assesses LLMs' ability to understand code semantics. Without reasoning abilities, it is hard to believe the generalizability of models to different and unseen contexts~\cite{du2023classeval,jimenez2023swe}. This demand has motivated a growing research effort to develop various code-reasoning evaluation tasks and benchmarks. 

Existing approaches~\cite{gu2024cruxeval,liu2024codemind, beger2025coconut, chen2024runtime} assess the code reasoning abilities of LLMs using LLM-generated, e.g., \crux~\cite{gu2024cruxeval}, or human-written programs, e.g., \avatar~\cite{ahmad2021avatar}, \ceval~\cite{du2023classeval}, and \heval~\cite{chen2021evaluating}. \crux contains simple, standalone Python programs. \avatar and \heval programs are algorithmically complex and potentially challenging to reason about, but far from representing real-world code repositories. Surprisingly, state-of-the-art LLMs, \emph{if} evaluated for code reasoning, mostly only consider very simple, LLM-generated \crux programs. The high performance of LLMs on reasoning tasks using \crux programs, alongside their struggles in real-world settings, suggests the need for better code-reasoning benchmarks.

\begin{wrapfigure}{r}{0.4\columnwidth}
  \vspace{-11pt}
  \includegraphics[width=0.4\columnwidth]{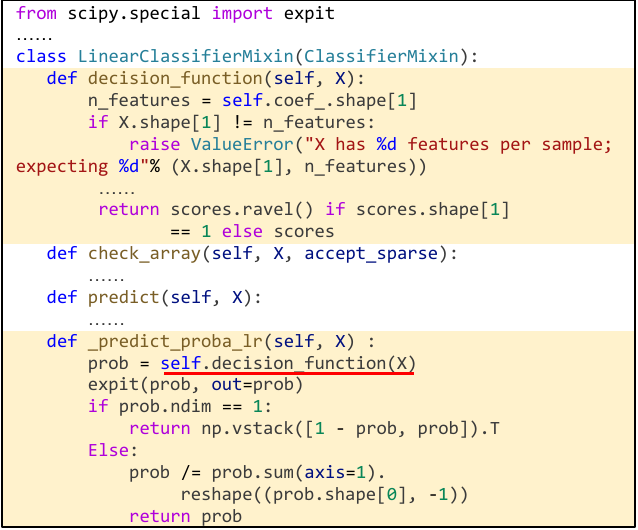}
  \vspace{-18pt}
  \caption{A real-world reasoning problem (highlighted methods)}
  \vspace{-12pt}
  \label{fig:reasoning-problem-example}
\end{wrapfigure}
For a systematic comparison of LLMs' code reasoning performance on real-world programs and those used in code reasoning studies, we constructed a dataset of \num{1200} \emph{reasoning problems} (details in \S \ref{sec:methodology}) from existing code reasoning evaluation benchmarks (\avatar, \ceval, \crux, and \heval), as well as real-world projects (from widely-used \SweB and freshly-mined \Real). The former provides LLM-generated and human-synthesized algorithmically complex programs, and the latter diversifies the dataset with real-world complexities. A reasoning problem is a method, passed along with its callees, to an LLM to predict its properties given an input or output. For simple programs such as those from \crux, the reasoning problem is identical to the program. For complex programs consisting of multiple methods or classes, the reasoning problem is a dynamic slice~\cite{agrawal1990dynamic}, i.e., a sequence of methods that will be directly or indirectly called during the execution of the method of interest. Figure~\ref{fig:reasoning-problem-example} shows a real-world reasoning problem, 
with \texttt{\small{\_predictproba\_lr}} as the method of interest: we ask LLM to predict input, output, loop variables, and branch decisions of it. \texttt{\small{\_predictproba\_lr}} directly invokes \texttt{\small{decision\_functions}}, which will be included as a necessary context to LLM. Other methods in the surrounding class are not part of the reasoning problem. We refer to the collected dataset as \name\footnote{\name stands for \underline{Re}alistic \underline{Re}soning \underline{bench}mark.}.  

\input{Tables/complexity}

We first analyze the complexity of \name problems from different perspectives using nine metrics presented in Table~\ref{metrics-complexity}\footnote{We make no claim to the completeness of these metrics. However, even within this set of metrics, we show that current reasoning problems lag behind real-world programs, motivating the inclusion of real-world problems to evaluate LLMs' code-reasoning abilities.}: Cyclomatic Complexity ($M_1$), Compound Predicates ($M_2$), Nested Constructs ($M_3$), Structural Complexity ($M_4$), Third-party APIs ($M_5$), Inter-class Dependency ($M_6$), Intra-class Dependency ($M_7$), Primitive Variable Count ($M_8$), and Complex Variable Count ($M_9$). $M_1$, $M_6$, and $M_7$ are classic code complexity metrics~\cite{mccabe1976complexity}. Prior studies have also shown that LLMs struggle with code reasoning when the program contains compound predicates, deeply nested constructs, and complex code structures such as list comprehensions, API calls, and compound variables~\cite{liu2024codemind,hooda2024large}. 
 
\begin{figure*}[t]
    \centering
    \includegraphics[width=0.99\textwidth]{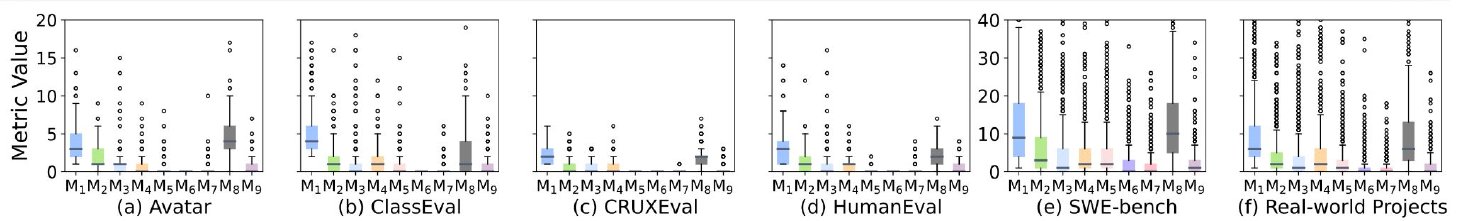}
    \vspace{-12pt}
    \caption{Complexity distribution of reasoning problems from prior techniques, compared to real-world programs. The metric value for sub-figures (a)--(d) ranges from $0$ to $20$, and for sub-figure (e)--(f) ranges from $0$ to $40$}
    \vspace{-10pt}
    \label{fig:benchmark-comparison}
\end{figure*}

Figure~\ref{fig:benchmark-comparison} illustrates the complexity distribution of collected reasoning problems per source of extraction. It is evident that reasoning problems from \real consistently show higher complexity across all nine selected metrics. Inter-class dependencies ($M_6$) do not exist in \heval, \avatar, \crux, and even \ceval. Moreover, reasoning problems in \crux, \emph{one of the most widely used benchmarks to assess code reasoning abilities of even the most recent powerful LLMs}, contain no third-party API usage ($M_5$) and near-zero intra-class dependencies ($M_7$). Such complexities exist in real-world programs and can challenge LLMs' code-reasoning abilities, motivating a study to evaluate and analyze LLMs' code reasoning under real-world complexity. 

We categorize \name reasoning problems into \emph{two} groups, \emph{Lower Complexity (LC)} and \emph{Higher Complexity (HC)}, via a principled majority-vote mechanism over the nine interpretable code complexity metrics. This categorization enables a more diagnostic evaluation of LLMs' code-reasoning abilities with a focus on real-world complexity. To ensure a rigorous and sound evaluation of LLMs under two categories and increase confidence in the generalizability of the claims, \name automatically tunes majority-voting thresholds and assesses the robust separation of the two complexity levels using Silhouette Analysis~\cite{shahapure2020cluster} (to determine and remove borderline reasoning problems) and Davies–Bouldin Index~\cite{davies2009cluster} calculation (to further assess the quality of category separation). Not surprisingly, all programs in prior code-reasoning studies belong to the \emph{LC} category, whereas more complex reasoning problems in the \emph{HC} category occur in real-world projects. 

The Real-world code contains various types of variables. Unlike prior studies that focus only on primitive and a few compound types and introduce them to LLMs as is, our pipeline leverages static and dynamic analysis to \emph{serialize} and \emph{deserialize} all objects, representing them in the prompt as a JSON structure. It also instructs the models to predict variable values through adaptive few-shot learning (\S \ref{subsec:prompt-strategies}), minimizing the impact of ambiguity and task difficulty on LLMs' performance~\cite{mann2020language}. Our pipeline is fully automated, with test execution to validate LLM predictions rather than a string-based comparison of predictions and ground truth used by most prior techniques. This helps \emph{automated detection of false negatives}, i.e., cases that differ from the collected ground-truth values but are correct with respect to program execution (\S \ref{subsub:false-negatives}). We also propose a series of new metrics to evaluate \emph{partial} code reasoning abilities of LLMs (\S \ref{subsec:eval-metrics}). 

We evaluate \emph{five} reasoning\footnote{Our terminology sources from model developers claiming these LLMs reason/think longer to respond compared to other model series by the same company~\cite{openai_gpt5mini_2025,openai_gpt5mini_2025,liu2025deepseek}.} LLMs 
and their \emph{five} non-reasoning/low-reasoning\footnote{“Non-reasoning” refers to (1) reasoning-capable LLMs with reasoning disabled~\cite{reasoning-tokens}, and (2) general-purpose LLMs that have not undergone reasoning-focused post-training. "Low-reasoning" refers to setting the reasoning budget/effort to "Low"~\cite{reasoning-budget}.} counterparts on \emph{four} code reasoning tasks, input, output, loop, and branch prediction,
using $500$ \name-lite reasoning problems (\S \ref{rq:effectiveness}). The extensive empirical study confirms that complex real-world code is substantially more challenging than the programs commonly used in prior evaluations: the average reasoning performance of the studied LLMs in input, output, loop, and branch prediction \emph{drastically drops} by $37.36\%$,  $36.16\%$, $20.90\%$, and $48.60\%$, respectively, when evaluating on \emph{LC} compared to \emph{HC} problems.
\name leverages static and dynamic analyses to analyze the impact of various programming constructs, call chain size, prompting/training strategies, overall program complexity, and individual complexity metrics on the reasoning abilities of LLMs:
\begin{itemize}[leftmargin=*]
    \item LLMs struggle the most with nested constructs in all tasks. Loops and conditional statements are particularly challenging for backward reasoning in predicting inputs (\S \ref{subsub:constrcuts}). 
    \item  The call chains are longer in the problems for which LLMs failed to predict their inputs or outputs correctly. The average call chain length for successful output prediction is longer than that for input prediction, demonstrating better abilities of LLMs in forward reasoning (\S \ref{subsub:call-chain}). 
    \item Models with high reasoning effort consistently outperform low-reasoning variants and general models on code reasoning tasks (\S \ref{subsub:training-strategies} and \S \ref{rq:additional}).
    \item A statistical analysis shows a moderate to strong negative correlation between LLMs' reasoning performance and individual code complexity metric value (\S \ref{rq:reasoning-failure}).
    
\end{itemize}

We finally follow a systematic approach to categorize reasoning failures into $18$ categories, providing deeper insight into what can be improved in the next generation of LLMs for better code reasoning (\S \ref{rq:reasoning-failure}). Our notable contributions are: (1) a dataset of reasoning problems containing real-world code, complex objects, long call chain slices (including intra- and inter-class dependencies), and third-party API invocation. Prior work either do not evaluate LLMs' code reasoning on real-world code, or only study real-world code without complex or compound types due to difficulty of dataset construction; (2) a systematic and adaptive approach to categorize collected reasoning problems into two well-separated and semantically meaningful complexity groups; (3) automated prompt crafting to evaluate four code reasoning tasks concerning complex variable types in real-world code; (4) proposing fine-grained evaluation metrics to measure reasoning success; (5) large-scale evaluation of code reasoning of ten general and reasoning LLMs under \name; and (6) a systematic categorization of reasoning failures into $18$ groups accompanied by case studies\footnote{Artifacts are publicly available~\cite{website}.}.

%% file: Tables/complexity.tex
\begin{wraptable}{t}{0.5\textwidth}
\centering
\footnotesize
\vspace{-12pt}
\caption{List of identified metrics.}
\vspace{-10pt}
\label{metrics-complexity}
\begin{adjustbox}{width=0.5\textwidth,center}
\begin{tabular}{|p{0.4cm}|p{2.9cm}|p{5.2cm}|} 
\hline
  \multicolumn{1}{|c|}{\textbf{ID}} &
  \multicolumn{1}{|c|}{\textbf{Metric}} &
  \multicolumn{1}{c|}{\textbf{Description}} \\ \hline

   $M_1$&
  Cyclomatic Complexity&
For a Control Flow Graph (CFG) of a program with $N$ nodes, $E$ edges, and $P$ connected components, $M_1 = E - N + 2 \times P$\\ \hline

$M_2$&
  Compound Predicates &
Number of compound predicates (sub-predicates connected with boolean, comparison, binary, or unary operators). \\ \hline

$M_3$&
  Nested Constructs &
    nested levels in recursive (for and while loops) and conditional constructs. \\ \hline

$M_4$&
  Structural Complexity &
    Number of complex code structures: list comprehension, dict comprehension, set comprehension, generator, lambda expression, list, thread, recursive function, and decorator. \\ \hline

$M_5$&
  Third-party APIs &
    Number of third-party API calls. \\ \hline

$M_6$&
  Inter-class Dependencies &
    Number of inter-class dependencies. \\ \hline

$M_7$&
  Intra-class Dependencies &
    Number of intra-class dependencies. \\ \hline

$M_8$&
  Primitive Variable Count&
    Number of primitive type variables. \\ \hline

$M_9$&
  Complex Variable Count&
    Number of complex type variables, e.g., dict. \\ \hline
\end{tabular}
\end{adjustbox}
\vspace{-10pt}
\end{wraptable}

%% file: Sections/Method.tex
\section{\name}
\label{sec:methodology}

Our systematic approach for benchmark construction includes three main steps of reasoning problem selection (\S \ref{subsec:dataset-collection}), variable value extraction (\S \ref{subsec:input-output}), and complexity level categorization (\S \ref{subsec:benchmark-construction}). 

\vspace{-8pt}
\subsection{Reasoning Problem Selection}
\label{subsec:dataset-collection}

\input{Tables/filtering}

\name constitutes reasoning problems from prior code reasoning studies (\avatar, \ceval, \crux, and \heval) and $15$ freshly-mined open-source programs\footnote{Followed the procedure by Allamanis et al.~\cite{allamanis2021self} in mining, we collected programs among the most downloaded Python repositories on GitHub with $\#stars \ge 500$, $\#forks \ge 100$, and a runnable green test suite. The detailed information about the mined repositories is available on the artifact website~\cite{website}. The average number of stars, forks, and downloads for collected projects is $22K$ (min=$685$,max=$86.6K$), $79K$ (min=$409$,max=$33.6K$), and $539.3M$ (min=$27.7M$,max=$1438.4M$).}. 
In addition, as many recent LLMs and agents are being evaluated on \SweB, we extracted reasoning problems from \SweB repositories to enable cross-analysis between code reasoning and other programming tasks\footnote{Such analysis performed by Liu et al.~\cite{liu2024codemind,liu2025assessing} belong to future work.}.
For \avatar, \crux, and \heval, we consider each program in the dataset as a reasoning problem, as they are standalone methods. For \ceval, each method in the class and its callees in the dynamic slice (executed by the provided unit tests) is a reasoning problem. Since \ceval offers multiple tests for each method, we randomly select one to have unique reasoning problems. This process provides \num{1450} \emph{initial} reasoning problems from these datasets.

For real-world projects, we extract the methods under test by tracing executions of the corresponding test suite, providing us with \num{8603} methods. 
We exclude duplicates, i.e., methods with identical inputs and outputs (filtering Rule 1), and remove the methods with no return statements or input arguments (filtering Rule 2). Finally, we exclude methods whose variable types have circular dependency (filtering Rule 3), as they result in a non-terminating serialization loop (details in \S \ref{subsec:input-output}). 
Table~\ref{tab:filtering} shows the number of final \emph{unique} methods along with their surrounding classes and the callees in the dynamic slices, forming \num{3129} \emph{initial} reasoning problems from \SweB and \real. For the reasoning problems that have more than one unique (input,output) pair, we randomly select one to prompt LLMs. 

\subsection{Variable Value Extraction}
\label{subsec:input-output}

\begin{figure*}[t]
    \centering
    \includegraphics[width=0.93\linewidth]{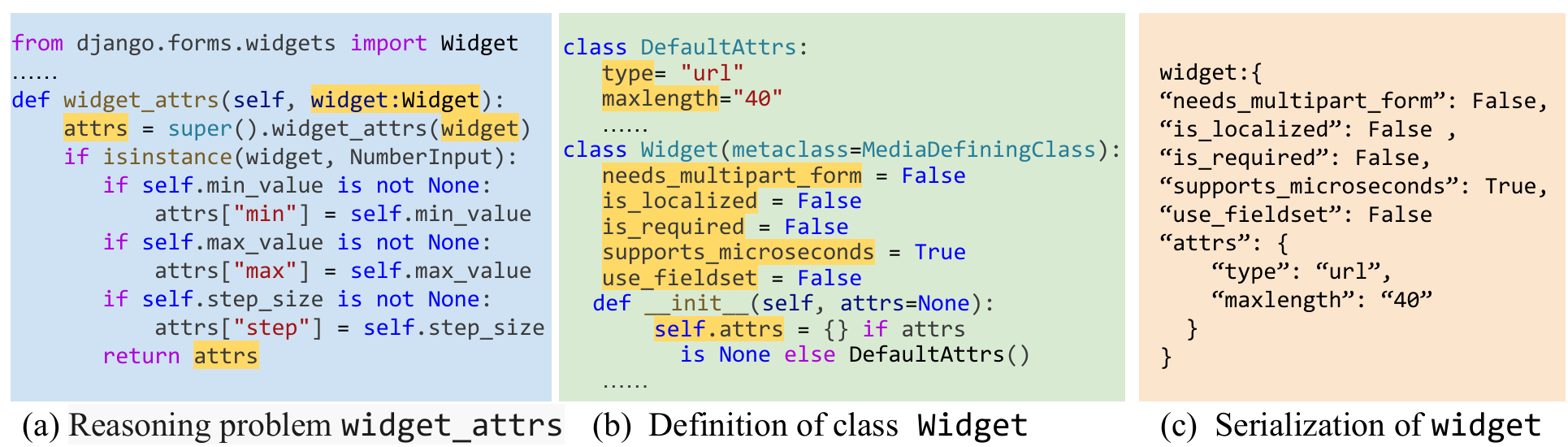}
    \vspace{-8pt}
    \caption{Example of custom types in real-world reasoning problems and \name serialization}
    \vspace{-15pt}
    \label{fig:complex-object}
\end{figure*}

\sloppy Existing datasets follow different designs for program execution. For \avatar, the inputs and outputs are identified in text files. For \heval and \crux, inputs and outputs are in the form of assert statements, e.g., in the \texttt{\small{\textcolor{Brown}{assert} \textcolor{Purple}{intersperse}(\textcolor{RoyalBlue}{[],7})==\textcolor{Green}{[]}}}, inputs are  \texttt{\small{\textcolor{RoyalBlue}{[]}}} and \texttt{\small{\textcolor{RoyalBlue}{7}}}, and output is \texttt{\small{\textcolor{Green}{[]}}}. For \ceval, \SweB, and real-world projects the inputs and outputs are not explicit but in the form of tests. \name implements a generic pipeline on top of all datasets, making it easy to extend in the future with problems from other mature datasets or evaluating other code reasoning tasks: given the main method, i.e., the one in the reasoning problem that we prompt LLMs to predict its properties, \name identities the input, output, and other variables through static analysis. It then runs existing tests and collects/serialize variable values at runtime. 

Many variables in \SweB and real-world reasoning problems are likely to be complex custom objects. Figure~\ref{fig:complex-object} (a) shows such an example, where the method \texttt{\small{widget\_attrs}} takes a \texttt{\small{Widget}} object as input. As shown in Figure~\ref{fig:complex-object} (b), this custom type has five \emph{class attributes} of \texttt{\small{bool}} type (\texttt{\small{needs\_multipart\_form}}, \texttt{\small{is\_localized}}, \texttt{\small{is\_required}}, \texttt{\small{supports\_microseconds}}, and \texttt{\small{use\_fieldset}}) and one \emph{instance attribute} of a custom type (\texttt{\small{DefaultAttrs}}). The \texttt{\small{DefaultAttrs}} also has two \emph{class attributes} of \texttt{\small{str}} type (\texttt{\small{type}} and \texttt{\small{maxlength}}). Variable \texttt{\small{widget}} initializes the return value \texttt{\small{attrs}}. 

\input{algorithm/variable_serializer}
Without a proper introduction of variable values to LLMs, they cannot reason how input values evolve to output (output prediction) or vice versa (input prediction). Also, unlike primitive types (e.g., \texttt{\small{int}} or \texttt{\small{bool}}) or core compound types such as \texttt{\small{list}} or \texttt{\small{set}}, whose corresponding values can be \emph{printed} during execution, custom objects may not implement \texttt{\small{str}} or \texttt{\small{repr}} methods by default; thereby, printing them provides memory address or hash value of the objects. To overcome this challenge, \name leverages program analysis to recursively decompose a complex type until it reaches primitive types, and stores them in a JSON format. When prompting for predicting variable values, \name represents variables given this structure, and instructs LLMs to understand the structure through in-context examples (\S \ref{subsec:prompt-strategies}). Algorithm~\ref{alg:serialization} shows an overview of this mechanism. Figure~\ref{fig:complex-object} (c) demonstrates the results of serializing the custom type of \texttt{\small{Widget}} to be printed during execution. 

\vspace{-8pt}
\subsection{Complexity Level Categorization}
\label{subsec:benchmark-construction}
\input{algorithm/categorization}

\name categorizes the reasoning problems into two groups: problems with \emph{Lower Complexity (LC)} and \emph{Higher Complexity (HC)}, using Algorithm~\ref{alg:cap}. The algorithm takes the \num{4579} ({\num{1450}}$+$\num{3129}) \emph{initial} reasoning problems from the previous step and returns a versioned code reasoning benchmark with balanced number of problems for each complexity level as output.

\begin{figure*}[t]
    \centering
    \includegraphics[width=\textwidth]{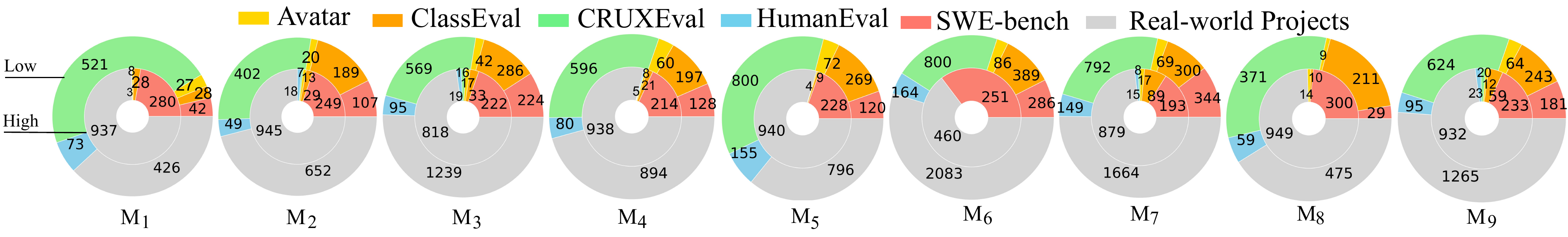}
    \vspace{-20pt}
    \caption{Breakdown of the reasoning problems across complexity levels and metrics for $T = 25\%$. The outer and inner donuts indicate the problems ($p_j$s) with their $l_i^j$ labeled as \lowComp (Low) and \highComp (High), respectively}
    \vspace{-15pt}
    \label{fig:donut-complexity}
\end{figure*}

For each reasoning problem $p_j$, the algorithm calculates corresponding complexity metric values (Table \ref{metrics-complexity}), yielding a vector, $[f_1^j, \cdots, f_{|CM|}^j]$ (Lines $3$--$4$). After calculating the complexity metric values for all the problems, it assigns each calculated value $f_i^j$ a label (\highComp or \lowComp), resulting in a vector of complexity labels for each problem, $[l_1^j, \cdots, l_{|CM|}^j]$ (Lines $5$--$14$): it sorts the problems according to $f_i^j$s (Lines $7$--$8$). The complexity label $l_i^j$ of the problems in the first $T$ percentile will be \lowComp (the $T\%$ of the programs with the lowest $f_i^j$s). The complexity label $l_i^j$ of the problems in the last $T$ percentile will be \highComp (the $T\%$ of the programs with the highest $f_i^j$s). The algorithm assigns no complexity label to the problems in between. Figure~\ref{fig:donut-complexity} shows the label breakdown of the initial reasoning problems across complexity metrics ($T=25\%$). 

After assigning the complexity labels, the algorithm categorizes the problems into two complexity levels following the majority voting principle (Lines $15$--$29$): $p_j$ is categorized under \textbf{Lower Complexity (LC)} or \textbf{Higher Complexity (HC)}, if the majority of its $f_i^j$s are labeled as \lowComp or \highComp, respectively. The \emph{majority} can be a number between $\lceil \frac{|CM|}{2} \rceil$\footnote{The majority voting $ \ge \lceil \frac{|CM|}{2} \rceil$ ensures that each problem is categorized under only one complexity level.} and $|CM|$. A lower majority will let more problems into each complexity category, while allowing more overlap between complexity dimensions. Given the purpose of \name, which aims to assess the reasoning performance shift concerning the complexity of the problems, well-separated and semantically meaningful categorization is essential. Algorithm~\ref{alg:cap} dynamically balances the trade-off between coverage and separability by iterating over the majority number (Lines $15$--$30$), forming the \emph{LC} and \emph{HC} categories (Lines $16$--$20$), applying Silhouette Analysis to determine and remove borderline problems (Line $21$), and evaluating the quality of the categorization using the Davies–Bouldin Index (Line $22$). If DBI indicates a semantically meaningful separation between the categories, the algorithm terminates (Lines $23$--$27$). Otherwise, it increments the majority for a better separation and repeats the categorization (Lines $28$--$30$). 

\begin{wrapfigure}{r}{0.45\columnwidth}
\vspace{-24pt}
  \includegraphics[width=0.45\columnwidth]{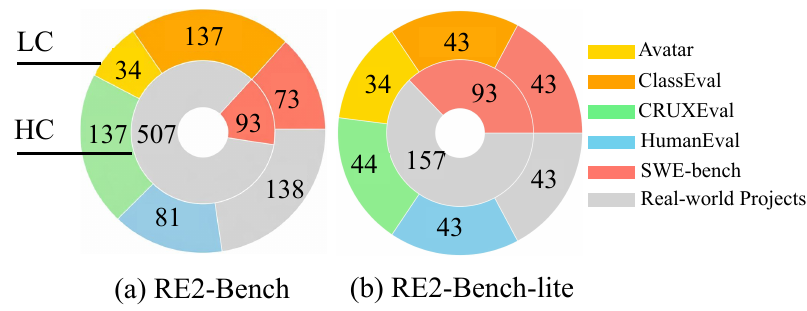}
  \vspace{-20pt}
\caption{\name and \name-lite overview}
\vspace{-10pt}
\label{fig:difficulty-level}
\end{wrapfigure}
The Silhouette score measures how well each reasoning problem fits within its complexity category, based on the distance to neighboring points within versus outside its cluster (\emph{local} reliability). DBI measures the average similarity between two complexity categories, based on intra-cluster dispersion and inter-cluster distance (\emph{global} soundness). Using both ensures that the categorization is not only locally precise but also globally valid, increasing the confidence of generalizability of claims. The rigor results in discarding a subset of initial reasoning problems that cannot be clearly assigned to two complexity categories. Algorithm~\ref{alg:cap} also balances the number of problems in \emph{LC} and \emph{HC} categories (Line $24$) to ensure fair  evaluation and comparison.

The final \name dataset constitutes \num{1200} reasoning problems\footnote{This achieved with $T = 25\%$ and a majority of six (out of nine). The Silhouette score for borderline detection is $0.15$.}, \num{600} under each complexity category (Figure~\ref{fig:difficulty-level}-a).
Notably, \emph{none of the problems used in prior work} make it to the \emph{HC} category. As we will show, evaluating LLM code reasoning on \emph{HC} problems results in a significant performance drop compared to \emph{LC} problems (\S \ref{rq:effectiveness}). Following existing work on benchmark construction, we also offer \name-lite, a lightweight subset of \name containing $500$ reasoning problems, $250$ \emph{LC} and $250$ \emph{HC}, shown in Figure~\ref{fig:difficulty-level}-b.

%% file: Tables/filtering.tex
\begin{wraptable}{r}{0.4\textwidth}
\vspace{-23pt}
\caption{Filtering results for extracting \SweB and \real reasoning problems under each rule.}
\vspace{-10pt}
\footnotesize
\label{tab:filtering}
\begin{adjustbox}{width=0.4\textwidth,center}
\begin{tabular}{|l|c|c|}
\hline
\multicolumn{1}{|c|}{} & \textbf{\# methods} & \textbf{\# (input,output)} \\ \hline
\textbf{Original}                    & \num{8603}                 & \num{35821083}          \\ \hline
\textbf{Rule 1}                      & \num{8603}                 & \num{421091}            \\ \hline
\textbf{Rule 2}                      & \num{4779}                 & \num{304298}            \\ \hline
\textbf{Rule 3}                      & \num{3129}                 & \num{193673}            \\ \hline
\end{tabular}
\end{adjustbox}
\vspace{-10pt}
\end{wraptable}

%% file: algorithm/variable_serializer.tex
\begin{wrapfigure}{r}{0.5\textwidth} 
  \vspace{-22pt}
  \begin{minipage}{0.5\textwidth}
    \begin{algorithm}[H]
      \caption{Variable Serialization}
      \label{alg:serialization}
      \footnotesize
      \begin{algorithmic}[1] 
        \Require Variable $V$
        \Ensure Serialized variable $V^\ast$
        \State $V^\ast \gets [\,]$  \Comment{initialize empty list}
        \If{\textproc{isCompound}$(V)\ \lor\ \textproc{isComplex}(V)$}
          \ForAll{$v_i \in V$}
            \If{\textproc{isPrimitive}$(v_i)$}
              \State $V^\ast.\textproc{append}(v_i)$
            \Else
              \State $V^\ast.\textproc{append}(\textproc{VariableSerialization}(v_i))$
            \EndIf
          \EndFor
        \Else
          \State $V^\ast.\textproc{append}(V)$
        \EndIf
        \State \Return $V^\ast$
      \end{algorithmic}
    \end{algorithm}
  \end{minipage}
  \vspace{-12pt}
\end{wrapfigure}

%% file: algorithm/categorization.tex
\begin{wrapfigure}{r}{0.5\textwidth} 
  \vspace{-20pt}
  \begin{minipage}{0.5\textwidth}
    \begin{algorithm}[H]
      \caption{Complexity Level Categorization}
      \label{alg:cap}
      \footnotesize
      \begin{algorithmic}[1] 
        \Require Reasoning Problems $P$, Cut-off Threshold $T$, Complexity Metrics $CM$
        \Ensure \name$_{version}$
        
        \State $P_{LC}, P_{HC}, \name_{version} \gets \emptyset$
        \State $TV_{low}, TV_{high} \gets [0]^{|CM|}$

        \ForAll{$p_j$ in $P$}
            \State $[f_1^j, \dots, f_{|CM|}^j] \gets $ \textproc{CalculateComplexity}($p_j, CM$)
        \EndFor
        
        \ForAll{$i \gets 1$ to $|CM|$}
            \State $SortedP \gets \emptyset$
            \ForAll{$p_j \in P$}
                \State $SortedP \gets$ \textproc{Sort}($SortedP, p_j, f_i^j$)
            \EndFor
            \State $TV_{low_i}$, $TV_{high_i}\gets$ \textproc{CutOffValuess}($SortedP,T$)
            \ForAll{$p_j \in P$}
                \If{$f_i^j \le TV_{low_i}$}
                    \State $l_i^j$ $\gets$ "L"
                \ElsIf{$f_i^j \ge TV_{high_i}$}
                    \State $l_i^j$ $\gets$ "H"
                \EndIf
            \EndFor
        \EndFor

         \ForAll{$majority \gets \lceil \frac{|CM|}{2} \rceil$ to $|CM|$}
            \ForAll{$p_j \in P$}
                \If{Count("H"$, [l_1^j, \dots, l_{|CM|}^j]$) $\ge majority$}
                    \State $P_{HC} \gets P_{HC} \cup p_j$
                \EndIf
                \If{Count("L"$, [l_1^j, \dots, l_{|CM|}^j]$) $\ge majority$}
                    \State $P_{LC} \gets P_{LC} \cup p_j$
                \EndIf
            \EndFor 
            
            \State \textproc{RemoveBorderlines}($P_{HC},P_{LC}$)
            \State $sep \gets$ \textproc{CheckClassification}($P_{HC},P_{LC}$)

            \If{$\lnot sep$}
                \State \textproc{Balance}($P_{HC},P_{LC}$)
                \State $\name_{version} \gets$ $P_{HC} \cup P_{LC}$
                \State $version \gets$ \textproc{SetSnapshot}($TV_{low}, TV_{high}$)
                \State $break$
            \Else
                \State $majority \gets majority+1$
                \State $continue$
            \EndIf
            
         \EndFor

        \State \Return $\name_{version}$
      \end{algorithmic}
    \end{algorithm}
  \end{minipage}
  \vspace{-25pt}
\end{wrapfigure}

%% file: Sections/Experiment.tex
\section{Empirical Evaluation}
\label{sec:evaluation}

Due to the cost of frontier reasoning models and to ensure a comprehensive evaluation from various aspects, we use \name-lite reasoning problems to investigate the following research questions:

\vspace{-3pt}
\begin{enumerate}[label=\textbf{RQ\arabic*:}]

    \item \textbf{Effectiveness.} To what extent do \name-lite problems under different complexity levels challenge the reasoning abilities of LLMs in widely-used input and output prediction tasks? What is the impact of false negatives?

    \item\textbf{Impacting Factors.} What factors make code reasoning more challenging for LLMs? 

    \item \textbf{Reasoning Failure Analysis.} What are the most common root causes of reasoning failures?

    \item \textbf{Ablation Study.} To what extent do different prompt crafting strategies contribute to the overall reasoning performance of LLMs?

    \item \textbf{Other Reasoning Tasks.} Do the findings for input and output prediction persist in loop and branch prediction tasks?

\end{enumerate}
\vspace{-3pt }

\input{Sections/3-1-setup}

\input{Sections/3-2-RQ1}

\input{Sections/3-3-RQ2}

\input{Sections/3-4-RQ3}
\input{Sections/3-6-RQ5}
\input{Sections/3-5-RQ4}

%% file: Sections/3-1-setup.tex
\noindent We evaluate \emph{five} reasoning LLMs (reasoning enabled or a high reasoning effort): \haiku~\cite{anthropic2025claudehaiku45systemcard}, \dsv~\cite{liu2025deepseek}, \geminip~\cite{google_gemini_3pro}, \gfmini~\cite{openai_gpt5mini_2025}, and \cwm~\cite{copet2025cwm}, and their \emph{five} non-reasoning (by design or by disabling the reasoning)/low-reasoning counterparts. This variety enables comparing the impact of reasoning-aware with general-purpose training. We set the temperature to $0$ for \haiku, \geminip, \dsv, and \cwm for reproducibility. \gfmini only runs with temperature $1$. As a result, our analysis will focus on comparing the performance of LLMs across different experimental settings, rather than comparing them with others. 
To handle cases where the LLM API returns an empty response due to service-side errors, we implement a fallback mechanism that reprompts the model and retries the request.

\subsection{Evaluation Tasks} 
\label{sub:tasks}
Input and output prediction are the most prevalent reasoning tasks used to evaluate the latest models. They directly test forward and backward end-to-end functional reasoning, thereby determining the \emph{upper bound} for LLMs' code-reasoning abilities. The first four research questions present the results for these two reasoning tasks, debunking the model's outstanding abilities in code reasoning using simple \crux programs. 
In addition, we evaluate subject LLMs on predicting a subset of runtime program properties, namely, loop and branch prediction. Loops and conditional branches identify the start or end of basic blocks; thereby, a correct or incorrect misprediction of them can reflect misprediction of program properties within the basic blocks~\cite{liu2025assessing}.

\vspace{-5pt}
\subsection{Prompting Strategies}
\label{subsec:prompt-strategies}

\begin{wrapfigure}{r}{0.51\columnwidth}
\vspace{-30pt}
  \includegraphics[width=0.51\columnwidth]{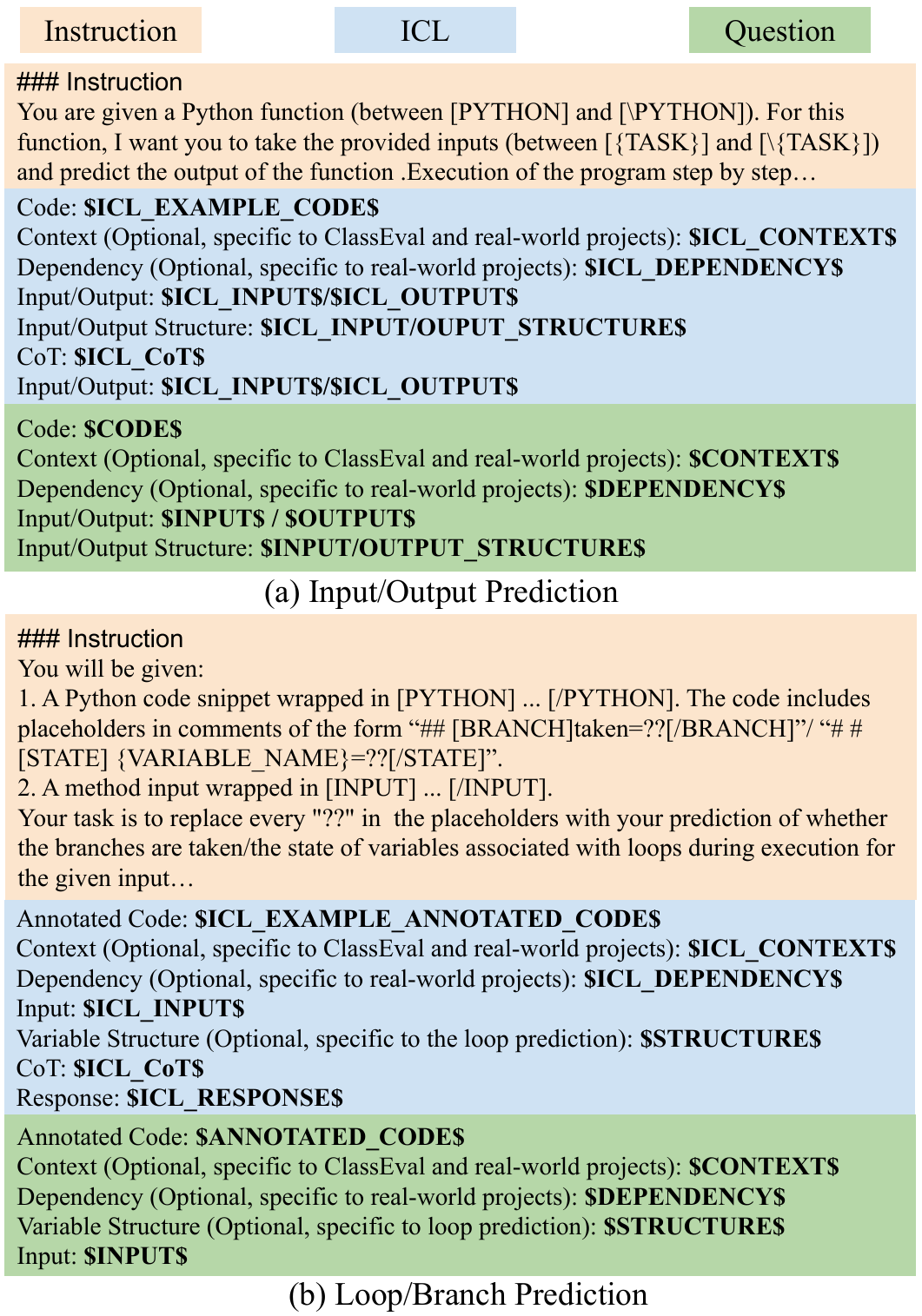}
  \vspace{-23pt}
\caption{Prompt template for Input and Output Prediction}
\vspace{-14pt}
\label{fig:prompt-template}
\end{wrapfigure}
We follow best practices to ensure proper prompt crafting for the studied tasks. Figure~\ref{fig:prompt-template} illustrates the prompt templates used for input, output, loop, and branch prediction \footnote{We have released our prompts on~\cite{website}.}. 
We use In-Context Learning (ICL) (\cite{mann2020language, ye2023compositional, dong2022survey}) examples to introduce the tasks to LLMs, and instruct them to perform step-by-step reasoning (\emph{explicit} reasoning similar to ICL example and \emph{implicit} reasoning in the instruction). 
Figure~\ref{fig:prompt-template}-a shows the prompts for input and output prediction.

Due to the complexity of inputs and outputs in real-world reasoning problems, we include additional \emph{hints} about \texttt{\small{Input/Output Structure}} in the prompt. Specifically, we provide the serialized format of the input or output to be predicted in the prompt, and ask LLMs to fill in the values instead of predicting the structure. For example, when we ask LLM to predict the input of the reasoning problem in Figure~\ref{fig:complex-object} (a) given the output, we provide the serialization of \texttt{\small{Widget}} object shown in Figure~\ref{fig:complex-object} (c) with all the values removed, so the LLM only predicts the values for each attribute. We include dependencies and class context in the prompt, if available. Our ablation study shows the effectiveness of different prompting strategies (\S \ref{rq:ablation}). 

Figure~\ref{fig:prompt-template}-b illustrates the prompt templates for loop prediction and branch prediction. Following best practices from Liu et al.~\cite{liu2025assessing}, we provide LLMs with the code annotated at loop or conditional statements. These annotations (prefixed with \texttt{"\#\#"}) indicate the properties to be predicted, and models are instructed to replace the placeholders marked as \texttt{"??"} with their answers. For loop prediction, the placeholders correspond to the states of loop variables and loop iterables\footnote{Please refer to \cite{liu2025assessing} for more information about loop variables, loop iterables, and the prompt construction}. For branch prediction, the placeholders require a binary decision indicating whether the branch is taken. As in input/output prediction, when loop variables or iterables have complex types, we include additional structural \emph{hints} in the prompt to specify the expected format of the prediction.

\vspace{-5pt}
\subsection{Evaluation Metrics}
\label{subsec:eval-metrics}

We measure code-reasoning abilities at two levels of granularity. At the highest level, the success of reasoning $RS_i$ is $1$ if the LLM correctly predicts \emph{all} the requested runtime properties, e.g., all the properties of a complex object or all the branches in the reasoning problem. Otherwise, $RS_i=0$ 
However, this metric is can be conservative and unfair. In the example of Figure~\ref{fig:complex-object}-c, if we have two LLMs, $LLM_x$ and $LLM_y$, correctly predicting \emph{one} and \emph{six} of the \emph{seven} properties of the input \texttt{\small{widget}}, both achieve $RS_i=0$. However, $LLM_y$ is superior to $LLM_x$ by predicting more properties. Similarly in loop and branch prediction tasks when a program contains multiple loops and branch statements, LLM should predict all the loop variable ($V_{l_j}$) and loop iterable ($I_{l_j}$) to achieve $RS_i=1$.

To promote a fairer evaluation of code reasoning under real-word settings, we introduce a metric to measure the \emph{partial} reasoning correctness, $RS_{partial_i}$, as below:

\begin{small}
\vspace{-5pt}
\begin{equation}
\label{ROP}
   RS_{partial_i}=\dfrac{\sum\limits_{j=1}^{m} \llbracket Pred_j = GT_j \rrbracket}{m} 
\end{equation}
\end{small}

For the input and output prediction tasks, $m$ is the total number of properties for each input or output variable to be predicted for problem $i$, $Pred_j$ is the LLM prediction for that property, and $GT_j$ is the ground truth value for the property. Using this metric, $LLM_x$ and $LLM_y$ achieve $RS_{partial_i}=0.14$ $(\frac{1}{7})$ and $RS_{partial_i}=0.86$ $(\frac{6}{7})$, respectively, providing a fairer comparison between models. 
For loop and branch prediction tasks, $m$ is the total number of loop and branch statements in problem $i$, $Pred_j$ is the LLM prediction for the loop or branch, and $GT_j$ is the corresponding ground truth value. For example, if a problem contains $5$ loops/branches and the model correctly predicts $3$ of them, then $RS_{partial_i}=0.6$ ($\frac{3}{5}$). For loop prediction, we consider a loop to be predicted correctly only if both loop variable and loop iterable are predicted correctly. Otherwise, partial reasoning may favor incoherence in reasoning~\cite{liu2025assessing}.

%% file: Sections/3-2-RQ1.tex
\subsection{RQ1: Effectiveness}
\label{rq:effectiveness}

\begin{figure*}
    \centering
    \includegraphics[width=0.99\linewidth]{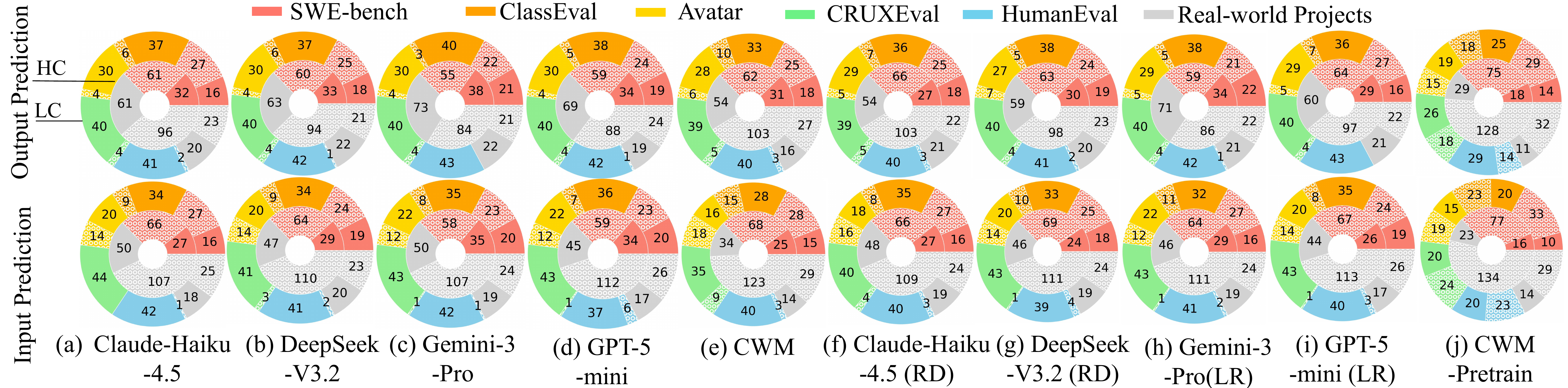}
    \vspace{-8pt}
    \caption{Distribution of the correct (solid) and incorrect (dotted) predictions ($RS$ metric) across complexity levels and datasets per task}
    \vspace{-8pt}
    \label{fig:succ-distribution}
\end{figure*}

For each reasoning problem, we compute both $RS$ and $RS_{partial}$ for \emph{LC} and \emph{HC} problems. Table~\ref{tab:main_results} presents the overall performance of the models on input/output prediction under these evaluation metrics, and Figure~\ref{fig:succ-distribution} shows the distribution of successful reasoning ($RS=1$) and failed reasoning ($RS=0$) across all the complexity levels in more detail. We report the breakdown per the origin or reasoning problems to further highlight the inflation of code reasoning abilities of LLMs when evaluated on \crux problems. 

\input{Tables/main_results}

\textbf{LLMs consistently achieve a higher reasoning success in \emph{LC} problems compared to \emph{HC} ones}: for input prediction, on average, $RS$ drops from $65.56\%$ to $28.20\%$, moving from \emph{LC} to \emph{HC} reasoning problems. For partial reasoning, $RS_{partial}$, the performance drops from $77.87\%$ to $60.77\%$, respectively. For output prediction, on average, $RS$ drops from $72.29\%$ to $36.13\%$, when moving from \emph{LC} to \emph{HC} reasoning problems. For partial reasoning, $RS_{partial}$, the performance drops from $73.73\%$ to $40.23\%$, respectively. 
\textbf{Reasoning-enabled LLMs outperform non-reasoning/low-reasoning LLMs} in both $RS$ and $RS_{partial}$, although marginally, by $4.82\%$ and $4.80\%$, respectively. This trend holds across both complexity groups: lowering or disabling reasoning effort leads to consistent declines in $RS$ and $RS_{partial}$. Among \emph{HC} problems, on average, the performance drop is $4.79\%$ and $4.92\%$, respectively. For \emph{LC} problems, on average, the drop is $4.87\%$ and $4.67\%$, respectively.

For all models and under all complexity groups, $RS_{partial}$ values are $22.44\%$ and $2.77\%$ higher than $RS$, on average, for input and output prediction. This indicates that \textbf{although LLMs have yet to achieve perfect code reasoning to predict all input or output values correctly, they can reason about some properties of inputs and outputs.} A strict metric, such as $RS$, can better reflect the end-to-end capabilities of the model in code reasoning. At the same time, there is a need for $RS_{partial}$ to reflect the incremental correctness or model's gradual progress~\cite{yu2024codereval,lyu2025top}. For the remainder of RQ1 and RQ2--RQ3 (\S \ref{rq:input-output-comparison}--\S \ref{rq:reasoning-failure}), we use $RS$ as it better reflects the overall LLMs' reasoning strength. We use $RS_{partial}$ during the ablation study (\S \ref{rq:ablation}). 

\begin{figure}[H]
    \centering
    \includegraphics[width=0.99\linewidth]{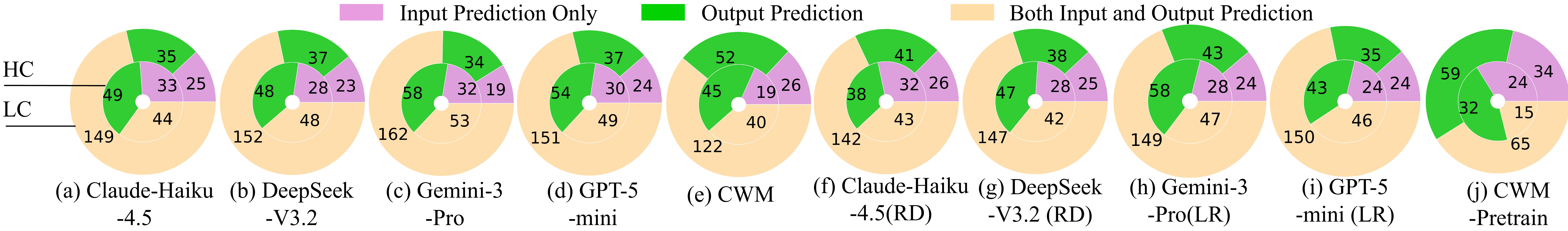}
    \vspace{-10pt}
    \caption{Unique and common problems each LLM succeeds in predicting their inputs and outputs.}
    \vspace{-10pt}
    \label{fig:successful_overlap}
\end{figure}

Finally, we analyze how the behavior of the same LLM changes from one reasoning task to another. Figure~\ref{fig:successful_overlap}, shows the number of unique and common problems each LLM succeeds in predicting their inputs and outputs. We observe that \textbf{LLMs are more successful in output prediction compared to input prediction, with an overall success rate of $54.21\%$ compared to $46.88\%$.} This is likely because input prediction is intuitively harder than output prediction, i.e., the LLM should perform only forward reasoning for output prediction, compared to both forward and backward reasoning for input prediction~\cite{ding2024semcoder}. Furthermore, as the complexity level of the reasoning problems increases, i.e., moving from outer doughnuts to inner ones, the percentages of overlap between input and output prediction success decrease. The gap is bigger for \emph{HC} problems, as real-world code contains more complex loops or conditional statements, making backward analysis considering complex objects as path conditions more difficult. 

From these results, it is evident that \textbf{the reasoning abilities of LLMs on \emph{LC} cannot generalize to \emph{HC} problems, the ones that LLMs need to deal with in practice.} This confirms the necessity of including reasoning problems with real-world difficulties in the dataset. Additionally, we argue that a comprehensive benchmark should include problems of different complexity levels, as \name does, due to the following reasons: (1) software engineering task complexities follow a long-tail distribution, i.e., some of them are small code with no notable complexity, and some are multiple methods or classes. Studies have shown that larger or better-trained LLMs may do worse on simple or trivial problems while improving on harder ones (inverse scaling)~\cite{mckenzie2023inverse,williams2024easy}; and (2) categorization enables regression testing across model versions or prompt revisions, which is specifically required for per-capability error analysis and targeted model diagnosis~\cite{ding2024easy2hard}.

\subsubsection{False Negatives in Input Prediction}
\label{subsub:false-negatives}

\begin{wrapfigure}{r}{0.5\columnwidth}
\vspace{-15pt}
  \includegraphics[width=0.5\columnwidth]{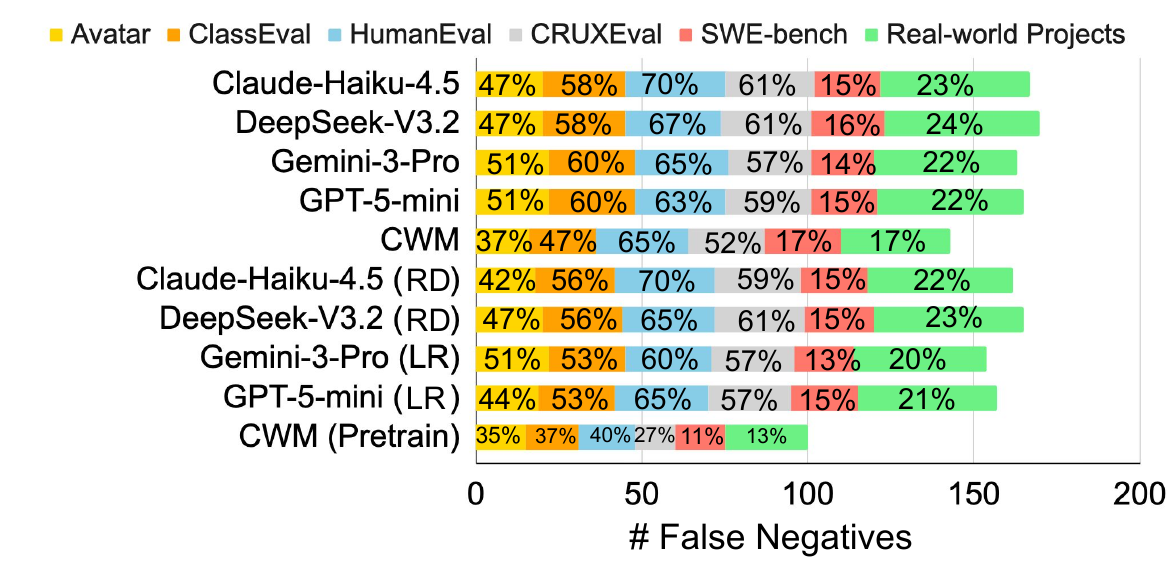}
  \vspace{-20pt}
\caption{False negative categorization per source}
\vspace{-10pt}
\label{fig:false-negative-stats}
\end{wrapfigure}
The possibility of false negatives in output prediction is narrow, since the outputs were either provided by the tests in the datasets, or we executed the tests to collect actual values during test execution. However, false negatives in input prediction are plausible: multiple inputs may result in the same output~\cite{dijkstra1982program,dinges2014targeted}. 
Our pipeline automates the detection of such cases for all problems (simple or real-world), and computes reasoning metrics accordingly: \textbf{all the reported numbers in the paper reflect the correction of false negatives to true positives}. To that end, it runs the standalone methods with the inputs predicted by LLMs and the ground truth input separately, and compares their execution results with assertions. For \ceval and real-world projects, an additional step of initializing the class with the predicted and ground truth  \texttt{\small{self}} arguments before executing methods is required. 

\begin{wrapfigure}{r}{0.38\columnwidth}
\vspace{-10pt}
  \includegraphics[width=0.38\columnwidth]{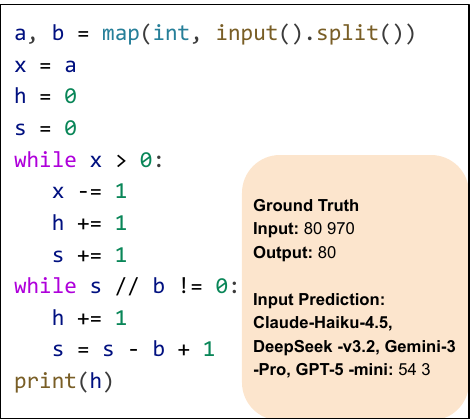}
  \vspace{-20pt}
\caption{False negative in \avatar~\cite{fn-avatar-geminip,fn-avatar-gpt,fn-avatar-dsv, fn-avatar-haiku}}
\vspace{-12pt}
\label{fig:false-negative-avatar}
\end{wrapfigure}

Figure~\ref{fig:false-negative-stats} demonstrates the number of false negatives for each LLM categorized per source of reasoning problems. We also report the ratio of false negatives to the total number of \name-lite problems from that source. Compared to the low-reasoning or non-reasoning counterparts, LLMs with higher reasoning effort yield more false negatives, which reflects their creativity and flexibility in code reasoning. We observe that \textbf{false negatives are more prevalent in \crux and \heval than in real-world problems}. This is likely because \crux and \heval consist of simple programs, making backward execution simulation less challenging for LLMs.

Figure~\ref{fig:false-negative-avatar} illustrates a false negative case from \avatar. The program reads two integers \texttt{\small{a}} and  \texttt{\small{b}}, initializes  \texttt{\small{h}} to $0$, and updates  \texttt{\small{h}} through a logic within two \texttt{while} loops. For the ground truth input \texttt{\small{"80 970"}}, the first loop executes exactly $80$ iterations (decreasing  \texttt{\small{x}} from $80$ to $0$), so both  \texttt{\small{h}} and  \texttt{\small{s}} become $80$. The second loop runs only if floor division of $s //b$ is zero, which holds for $80 // 970$, and execution does not enter the second loop. The program prints $80$ as the output. \haiku, \dsv, \geminip, and \gfmini predict \texttt{\small{"54 3"}} as the input. Although not equal to the ground truth input \texttt{\small{"90 970"}}, execution of program with this input results in the same output: the first loop iterates for $54$ times and updates \texttt{\small{h}} to $54$, and the second loop will run another $26$ iterations and finally update \texttt{\small{h}} to $80$.

Figure~\ref{fig:false-negative-swe} illustrates another example from a real-world project: the method \texttt{\small{split\_super\_sub}} takes \texttt{\small{text}} as input. The loop partitions the input into the base name, superscript (\texttt{\small{supers}}), and subscript (\texttt{\small{subs}}). After the loop, an if-condition checks whether the name ends with digits. If so, the number is placed at the beginning of the subscript. In this example, \geminip and \gfmini predict {\small\texttt{"alpha1\string^+"}}, whereas \dsv predicts \texttt{\small{"alpha\string_1\string^+"}}. Both predictions, however, yield the same output as the ground truth input \texttt{\small{"alpha\string^+\string_1"}}. All detected false negatives are available in our artifact website~\cite{website}.

%% file: Tables/main_results.tex
\begin{table*}[]
\setlength{\tabcolsep}{1pt}
\footnotesize
\caption{Reasoning performance of LLMs $\langle RS,RS_{partial} \rangle$ on input/output prediction using \name-lite. $RD$ refers to "reasoning disabled" and $LR$ refers to "low reasoning effort".}
\vspace{-10pt}
\label{tab:main_results}
\resizebox{\textwidth}{!}{
\begin{tabular}{|l|lll|lll|}
\hline
\multirow{2}{*}{\textbf{Subject LLMs}} &
  \multicolumn{3}{c|}{\textbf{Input Prediction\%}} &
  \multicolumn{3}{c|}{\textbf{Output Prediction\%}} \\ \cline{2-7} 
 &
  \multicolumn{1}{c|}{\textbf{HC}} &
  \multicolumn{1}{c|}{\textbf{LC}} &
  \multicolumn{1}{c|}{\textbf{Total}} &
  \multicolumn{1}{c|}{\textbf{HC}} &
  \multicolumn{1}{c|}{\textbf{LC}} &
  \multicolumn{1}{c|}{\textbf{Total}} \\ \hline
\textbf{Claude-Haiku-4.5} &
  \multicolumn{1}{l|}{$\langle 30.80, 64.00\rangle$} &
  \multicolumn{1}{l|}{$\langle 69.60, 81.65\rangle$} &
  $\langle 50.20, 72.83\rangle$ &
  \multicolumn{1}{l|}{$\langle 37.20, 41.42\rangle$} &
  \multicolumn{1}{l|}{$\langle 73.60, 75.33\rangle$} &
  $\langle 55.40, 58.38\rangle$ \\ \hline
\textbf{DeepSeek-V3.2} &
  \multicolumn{1}{l|}{$\langle 30.40, 63.16 \rangle$} &
  \multicolumn{1}{l|}{$\langle 70.00, 81.65\rangle$} &
  $\langle 50.20, 72.30\rangle$ &
  \multicolumn{1}{l|}{$\langle 38.40, 44.25\rangle$} &
  \multicolumn{1}{l|}{$\langle 75.60, 76.96\rangle$} &
  $\langle 57.00, 60.61 \rangle$ \\ \hline
\textbf{Gemini-3-Pro} &
  \multicolumn{1}{l|}{$\langle 34.00, 65.70\rangle$} &
  \multicolumn{1}{l|}{$\langle 72.40, 84.23\rangle$} &
  $\langle 53.20, 74.97 \rangle$ &
  \multicolumn{1}{l|}{$\langle 44.40, 49.21 \rangle$} &
  \multicolumn{1}{l|}{$\langle 78.40, 79.94\rangle$} &
  $\langle 61.40, 64.57\rangle$ \\ \hline
\textbf{GPT-5-mini} &
  \multicolumn{1}{l|}{$\langle 31.60, 64.87\rangle$} &
  \multicolumn{1}{l|}{$\langle 70.00, 81.77\rangle$} &
  $\langle 50.80, 73.32\rangle$ &
  \multicolumn{1}{l|}{$\langle 41.20 , 45.19\rangle$} &
  \multicolumn{1}{l|}{$\langle 75.20, 76.21\rangle$} &
  $\langle 58.20 , 60.70\rangle$ \\ \hline
\textbf{CWM} &
  \multicolumn{1}{l|}{$\langle 23.60, 55.14\rangle$} &
  \multicolumn{1}{l|}{$\langle 59.20, 73.10\rangle$} &
  $\langle 41.20, 64.12 \rangle$ &
  \multicolumn{1}{l|}{$\langle 34.00, 36.67\rangle$} &
  \multicolumn{1}{l|}{$\langle 69.60, 70.71\rangle$} &
  $\langle 51.80, 53.69\rangle$ \\ \hline
\textbf{Claude-Haiku-4.5 (RD)} &
  \multicolumn{1}{l|}{$\langle 30.00, 63.27\rangle$} &
  \multicolumn{1}{l|}{$\langle 67.20, 80.35\rangle$} &
  $\langle 48.60, 71.81\rangle$ &
  \multicolumn{1}{l|}{$\langle 32.40, 35.47\rangle$} &
  \multicolumn{1}{l|}{$\langle 73.20,74.65\rangle$} &
  $\langle 52.80, 55.06\rangle$ \\ \hline 
  
  \textbf{DeepSeek-V3.2 (RD)} &
  \multicolumn{1}{l|}{$\langle 28.00, 62.58\rangle$} &
  \multicolumn{1}{l|}{$\langle 68.80, 80.55\rangle$} &
  $\langle 48.40, 71.65\rangle$ &
  \multicolumn{1}{l|}{$\langle 35.60, 40.33\rangle$} &
  \multicolumn{1}{l|}{$\langle 74.00, 75.39\rangle$} &
  $\langle 54.80, 57.86\rangle$ \\ \hline

\textbf{Gemini-3-Pro (LR)} &
  \multicolumn{1}{l|}{$\langle 30.00, 63.73 \rangle$} &
  \multicolumn{1}{l|}{$\langle 69.20, 81.66\rangle$} &
  $\langle 49.60, 72.70\rangle$ &
  \multicolumn{1}{l|}{$\langle 42.00, 46.94\rangle$} &
  \multicolumn{1}{l|}{$\langle 76.80, 78.52\rangle$} &
  $\langle 59.40, 62.73\rangle$ \\ \hline

\textbf{GPT-5-mini (LR)} &
  \multicolumn{1}{l|}{$\langle 28.00, 63.58\rangle$} &
  \multicolumn{1}{l|}{$\langle 69.60, 81.44\rangle$} &
  $\langle 48.80, 72.51\rangle$ &
  \multicolumn{1}{l|}{$\langle 35.60, 39.87\rangle$} &
  \multicolumn{1}{l|}{$\langle 74.00, 75.19\rangle$} &
  $\langle 54.80, 57.53\rangle$ \\ \hline 
  \textbf{CWM (Pre-train)} &
  \multicolumn{1}{l|}{$\langle 15.60, 41.67\rangle$} &
  \multicolumn{1}{l|}{$\langle 39.60, 52.54\rangle$} &
  $\langle 27.60, 47.10\rangle$ &
  \multicolumn{1}{l|}{$\langle 20.50, 22.93\rangle$} &
  \multicolumn{1}{l|}{$\langle 52.50, 54.41\rangle$} &
  $\langle 36.50, 38.67\rangle$ \\ \hline \hline

  \textbf{Average} &
  \multicolumn{1}{l|}{$\langle 28.20, 60.77\rangle$} &
  \multicolumn{1}{l|}{$\langle 65.56, 77.87\rangle$} &
  $\langle 46.88, 69.32\rangle$ &
  \multicolumn{1}{l|}{$\langle 36.13, 40.23\rangle$} &
  \multicolumn{1}{l|}{$\langle 72.29, 73.73\rangle$} &
  $\langle 54.21, 56.98\rangle$ \\
  
  \hline
\end{tabular}
}
\vspace{-10pt}
\end{table*}

%% file: Sections/3-3-RQ2.tex
\subsection{RQ2: Impacting Factors}
\label{rq:input-output-comparison}

\begin{figure*}
    \centering
    \includegraphics[width=0.73\linewidth]{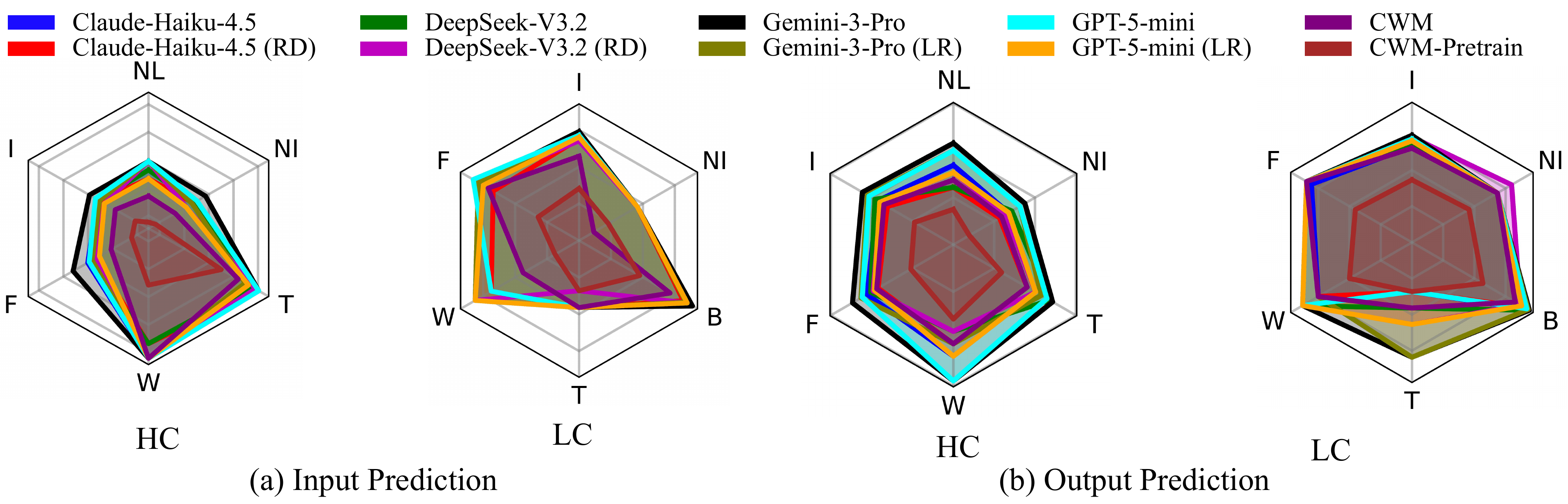}
    \vspace{-8pt}
    \caption{Impact of programming constructs on input and output prediction. Abbreviations are: Basic (B), For loop (F), If statement (I), Nested If (NI), Nested Loop (NL), Try statement (T), and While loop (W)}
    \vspace{-10pt}
    \label{fig:constructs}
\end{figure*}

\subsubsection{Program Constructs}
\label{subsub:constrcuts}

\begin{wrapfigure}{r}{0.55\columnwidth}
\vspace{-40pt}
  \includegraphics[width=0.55\columnwidth]{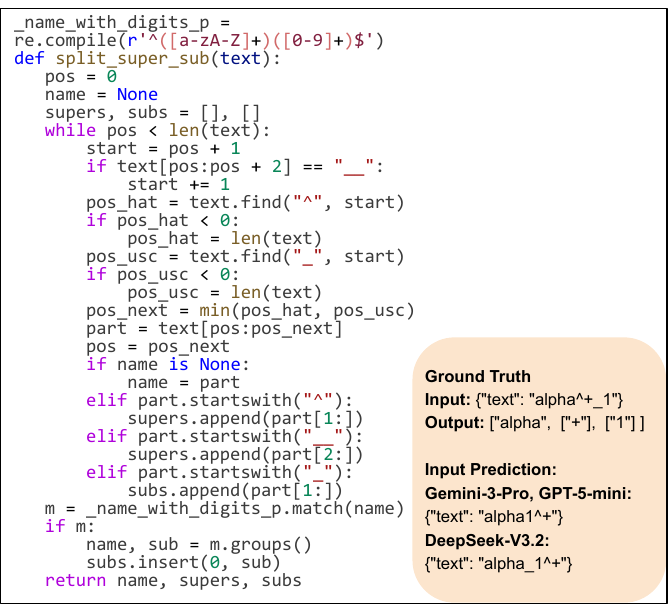}
  \vspace{-20pt}
\caption{False negative in real-world problems~\cite{fn-real-geminip, fn-real-gpt5, fn-real-dsv}}
\vspace{-10pt}
\label{fig:false-negative-swe}
\end{wrapfigure}
Complex program constructs can pose greater reasoning challenges for LLMs~\cite{liu2024codemind,hooda2024large}. To analyze the syntactic properties of the problems and their impact on the input and output reasoning, we used ExerScope~\cite{liu2025tool}. ExerScope uses static analysis to identify different programming constructs in Python code and label them with all existing code constructs. The existing labels implemented in ExerScope are \textbf{Basic (B)}---no specific programming constructs, \textbf{For (F)}---\texttt{\small{for}} loop, \textbf{If (I)}---\texttt{\small{if}} statement,  \textbf{Nested If (NI)}---nested \texttt{\small{if}} statements, \textbf{Nested Loop (NL)}---nested \texttt{\small{for}}, \textbf{Try (T)}--exception handling, and \textbf{While (W)}---\texttt{\small{while}} loop. After labeling the problems, ExerScope clusters them according to labels and visualizes the results. 

Figure~\ref{fig:constructs} shows the result of analysis per complexity level for input prediction (Figure~\ref{fig:constructs}-a) and output prediction (Figure~\ref{fig:constructs}-b). We observe that \textbf{LLMs struggle the most with nested constructs (\textbf{NI} and \textbf{NL} labels) in both input and output predictions.} Besides, loops and conditional statements are more challenging for backward reasoning to predict inputs, compared to straight-line programs. As a result, in RQ5 and for loop and branch prediction tasks, we evaluate the forward reasoning of LLMs concerning loops and branches (provide them with inputs). 

\subsubsection{Call Chain Size}
\label{subsub:call-chain}

The majority of the reasoning problems from real-world projects contain intra- and inter-class dependencies. As a result, for the prediction of input and outputs of the main method, LLMs need to reason about the control and data flow of methods called inside it. To further investigate the impact of call chain size on the reasoning performance of LLMs, we measured the call chain size, i.e., the number of methods involved in the reasoning problem slice. In the example of Figure~\ref{fig:reasoning-problem-example}, the call chain size is two, since the main method invokes only one other method.

\begin{figure*}
    \centering
    \includegraphics[width=0.99\linewidth]{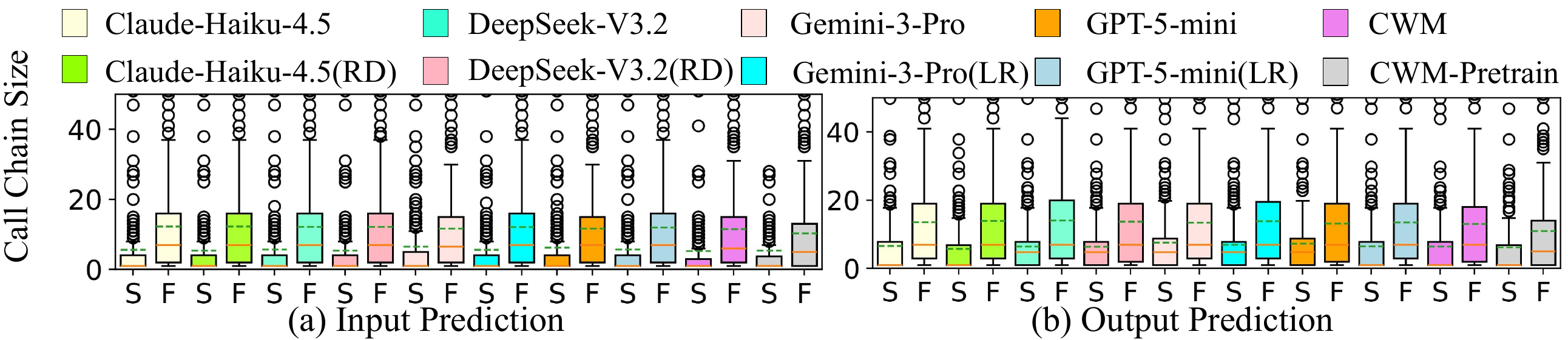}
    \vspace{-8pt}
    \caption{Comparison of the call chain size distribution between successful (\emph{S}) and failed (\emph{F}) reasoning}
    \label{fig:call-chain-size}
    \vspace{-10pt}
\end{figure*}

Figure~\ref{fig:call-chain-size} plots the range of call chain size for reasoning problems that each LLM succeeded (labeled as \emph{S}) or failed (labeled as \emph{F}) in predicting its inputs or outputs.  \textbf{The call chain size, on average (dashed green lines), is higher for the reasoning problems for which LLMs failed to predict their inputs or outputs correctly}. The average call chain size for successful output prediction is also higher than input prediction for each model, likely because forward reasoning is easier for LLMs than backward reasoning, specifically when programs become more complex.

\subsubsection{Training Strategies and Reasoning Efforts}
\label{subsub:training-strategies}

Table~\ref {tab:main_results} shows that reasoning LLMs (with high reasoning efforts) surpass their low-reasoning or non-reasoning versions in input and output prediction. Reasoning LLMs benefit from additional supervised fine-tuning and reinforcement learning that enhances their reasoning overall, and our results show that reasoning ability generalizes to code reasoning. For a deeper analysis, we looked into successful, unique, and overlapping reasoning problems in each model's family, concerning both input and output prediction. 

\begin{figure}[H]
\centering
\vspace{-3pt}
  \includegraphics[width=0.95\textwidth]{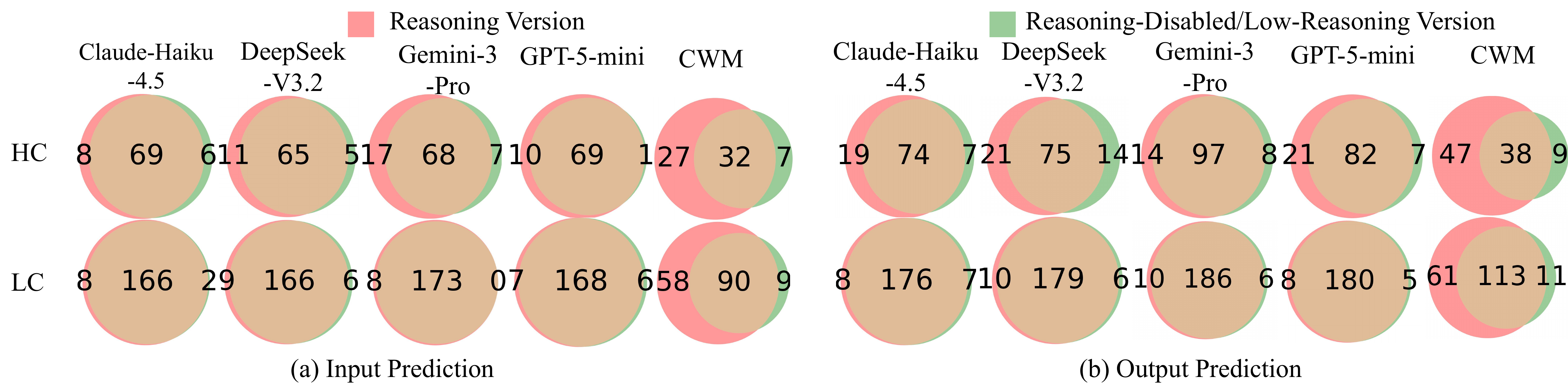}
  \vspace{-10pt}
\caption{Comparison between the performance of reasoning and low-reasoning/non-reasoning LLMs}
\vspace{-8pt}
\label{fig:reasoning-general-comparison}
\end{figure}

Figure~\ref{fig:reasoning-general-comparison} shows the comparison between different reasoning efforts in each family (columns) for each complexity category (rows), separated for input and output prediction. Overall, \textbf{models with high reasoning efforts can reason about more unique problems compared to low-/non-reasoning LLMs}. Despite the overall superiority, we can observe that there are $129$ cases where low-/non-reasoning LLMs correctly reason about inputs or outputs, but their counterpart reasoning LLM fails. We speculate this is due to \emph{Inverse Scaling in Test-Time Compute}, i.e., enabling higher reasoning effort or longer chains of thought can sometimes degrade accuracy, causing the model to overthink and arrive at an incorrect answer even when a lower-effort setting succeeds~\cite{gema2025inverse}.

%% file: Sections/3-4-RQ3.tex
\subsection{RQ3: Reasoning Failure Analysis.}
\label{rq:reasoning-failure}

\input{Tables/correlation}

We follow a \emph{semi-automated} approach for a \emph{systematic} root-cause analysis of reasoning failures. To that end, we revisit the complexity metrics (Table~\ref{metrics-complexity} that were used to motivate the need for better benchmarks. We first use the \emph{point biserial correlation}~\cite{kornbrot2014point} ($r_{pb}$) to measure the strength and direction of the relationship between reasoning results (binary $RS_i$ values) and complexity metric values (continuous values of $M_1$ to $M_9$) across all the reasoning problems. The coefficient $r_{pb}$ takes a value between $-1$ and $1$. The sign indicates positive or negative correlation, and the value determines the strength of correlation. Values close to zero indicate little to no relationship. Table~\ref{tab:correlation-analysis} shows the results of this analysis. Overall, there is always a strong to moderate negative correlation between the success of LLMs in studied code reasoning tasks and measured complexity metrics. 

\begin{figure}[H]
    \centering
    \vspace{-10pt}
    \includegraphics[width=\linewidth]{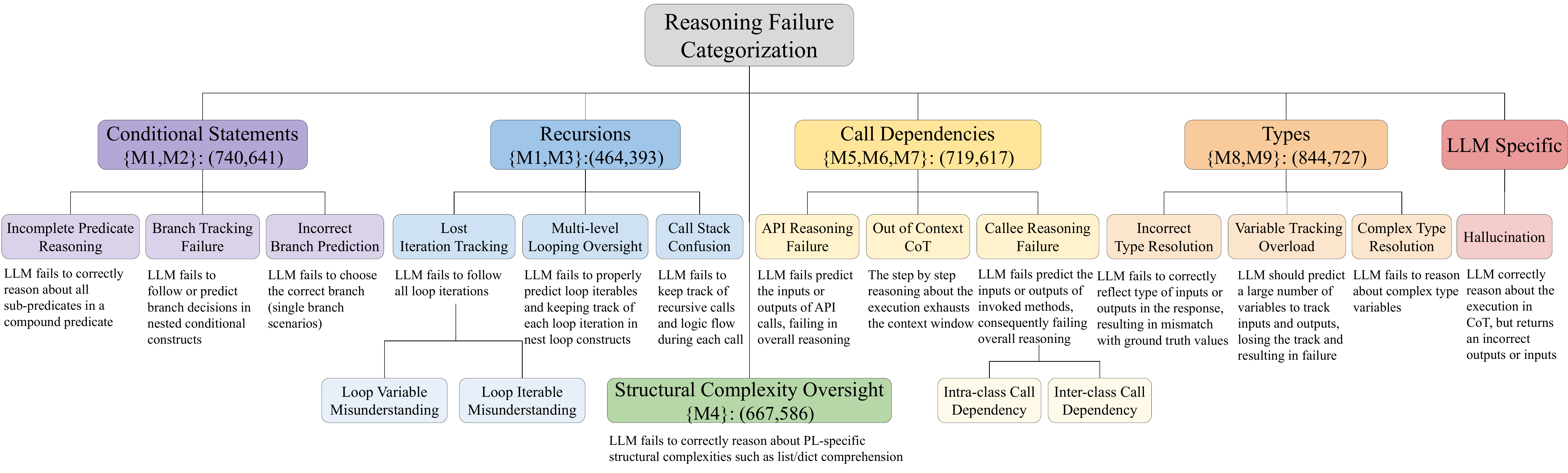}
    \vspace{-20pt}
    \caption{Categorization of reasoning failures}
    \vspace{-15pt}
    \label{fig:taxonomy}
\end{figure}
\begin{figure}[H]
    \centering
    \includegraphics[width=0.98\linewidth]{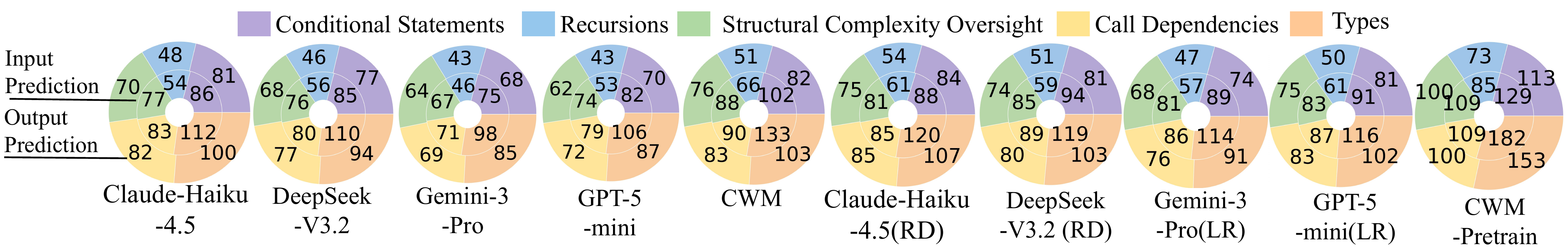}
    \vspace{-10pt}
    \caption{Breakdown of the reasoning failure categorization per individual LLM}
    \vspace{-10pt}
    \label{fig:LLM-taxonomy-breakdwon}
\end{figure}

For root cause analysis, we start by creating reasoning failure categories corresponding to individual or a group of complexity metrics, as shown in Figure~\ref{fig:taxonomy}: \textbf{Conditional Statements} (purple) with \emph{$M_1$, and $M_2$}, \textbf{Recursions} (blue) with \emph{$M_1$, and $M_3$}, \textbf{Structural Dependency Oversight} (Green) with \emph{$M_4$}, \textbf{Call Dependencies} (Yellow) with \emph{$M_5$, $M_6$, and $M_7$}, and \textbf{Types} (Orange) with \emph{$M_8$, and $M_9$}. Next, we assign reasoning failures into these categories as follows: a reasoning failure belongs to cluster $i$ if its $M_i$ metric is labeled as \emph{H (High)}, allowing failures to be grouped into one or more clusters. For example, a reasoning failure with \emph{High} label for its $M_1$, $M_2$, and $M_3$ will be categorized under both \emph{Conditional Statements} and \emph{Recursions} categories. 
The duplication under different categories allows for incorporating different possibilities for failures. The number of reasoning failures corresponding to each category is reported in Figure~\ref{fig:taxonomy}: the tuple under the category name shows the number of reasoning failures for input prediction and output prediction, accumulated for all LLMs. For example, there are $921$ and $811$ reasoning failures under \emph{Conditional Statements} category. Figure~\ref{fig:LLM-taxonomy-breakdwon} also shows the breakdown of the numbers for each LLM. 


Once reasoning failures are assigned to categories, we randomly selected $40$ instances from each cluster for each task to manually analyze the LLM CoT traces, providing us with $400$ reasoning failure instances ($200$ for input prediction and $200$ for output prediction). Selection of $40$ instances allows us to detect any theme appearing in $\ge10\%$ of reasoning failures at each category with $95\%$ confidence. In the random selection process, we selected an equal number of \emph{five} reasoning failure instances for each category for a given LLM. As a result of this manual investigation, we derive reasoning failure subcategories, e.g., Call Stack Confusion, Variable Tracking Overload in Figure~\ref{fig:taxonomy}. Through this process, we also identified cases that could have been categorized under a unique group, \emph{LLM Specific} (Red) category in Figure~\ref{fig:taxonomy}. While we do not claim completeness due to incorporating sampling (although with a high confidence interval) and studying two reasoning tasks, the multi-step process, grounded in statistical analysis and manual investigation, is systematic and rigorous, which can be extended/generalized to other reasoning tasks. 
\begin{figure*}[t]
    \centering
    \includegraphics[width=0.95\linewidth]{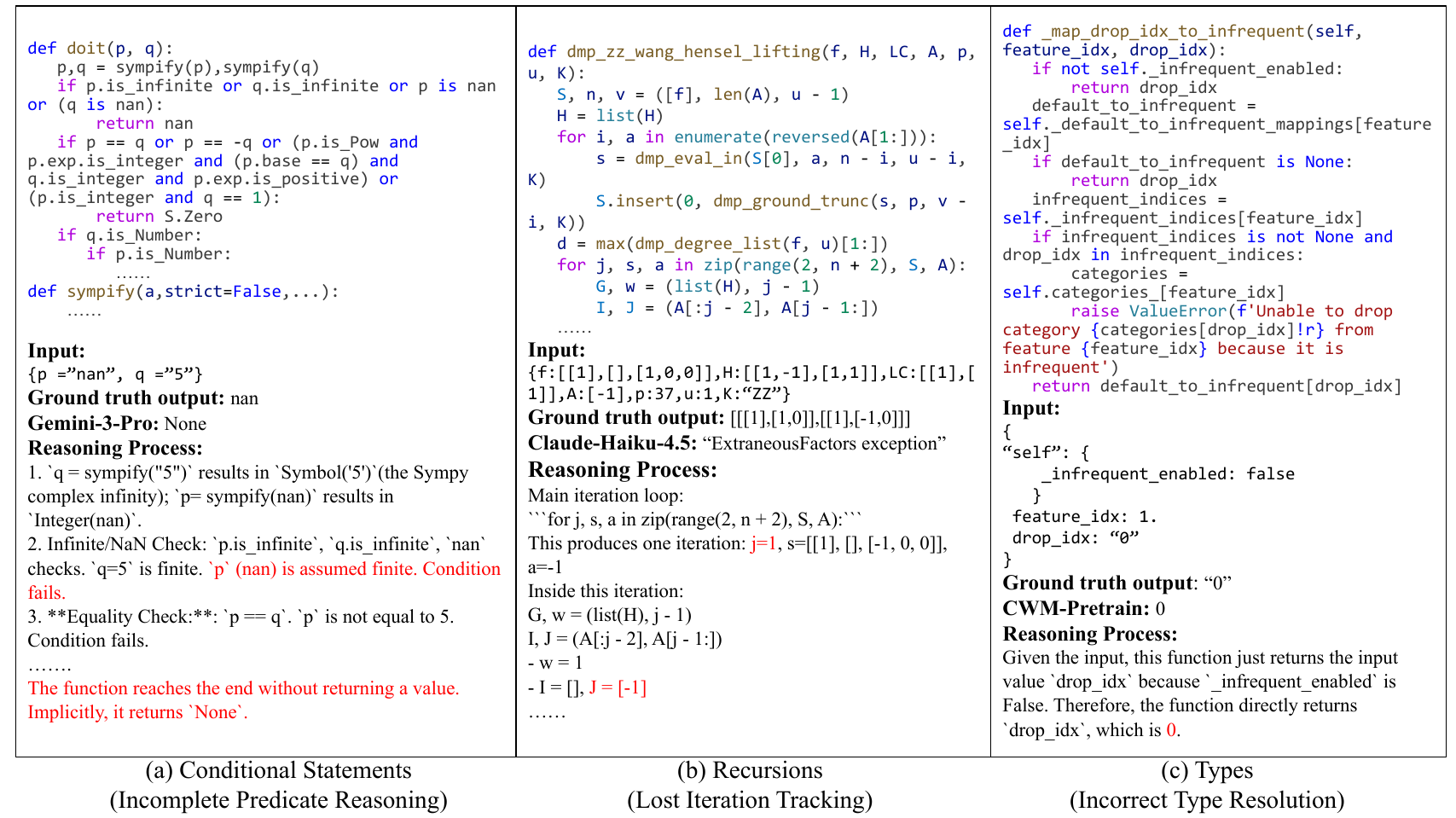}
    \vspace{-8pt}
    \caption{Examples of reasoning failures and their failure categories~\cite{failure-conditional,failure-recursion,failure-type}}
    \vspace{-15pt}
    \label{fig:taxonomy-examples}
\end{figure*}

\subsubsection{Qualitative Analysis}
We discuss a few examples of reasoning failures identified from previous analysis\footnote{Additional examples are available in our artifact website~\cite{website}.}. Figure~\ref{fig:taxonomy-examples}-a shows an example where \geminip predicts an incorrect output for
a real-world problem due to incorrect reasoning about conditional properties. The code takes two strings (\texttt{\small{p="nan"}} and \texttt{\small{q="5"}}) as inputs.
Since \texttt{\small{p}} satisfies \texttt{\small{p is nan}}, the code takes the first \texttt{\small{if}} branch and returns \texttt{\small{nan}}. When asked to predict the output, \geminip incorrectly assumes a \texttt{\small{nan}} value is \emph{finite} and the first predicate to be \texttt{\small{False}}, and skips evaluation of the other two sub-predicates in the first \texttt{\small{if}} statement, predicting the branch is not taken. It also evaluates other branch decisions to \texttt{\small{False}} and the final output to be \texttt{\small{None}}. 

Figure~\ref{fig:taxonomy-examples}-b presents an incorrect output prediction by \haiku on a real-world problem, due to an incorrect simulation of loop execution. The model correctly reasons about the number of iterations of the second \texttt{\small{for}} loop, but incorrectly predicts \texttt{\small{j}} value in the first iteration.
This mistake propagates to the state of \texttt{\small{J}} and the model consequently raises an \texttt{\small{ExtraneousFactors exception}} in the execution simulation. LLMs may also successfully reason about variable values, but not types. 
In the real-world examples of Figure~\ref{fig:taxonomy-examples}-c, \cwm-Pretrain correctly reasons about the conditional statements and takes the first branch in its reasoning. However, it converts the string input \texttt{\small{"0"}} of \texttt{\small{drop\_idx}} into the integer value $0$ in its predicted output. Although the predicted value is correct, the type is not, resulting in a test failure by our pipeline, and is considered an incorrect prediction.




%% file: Tables/correlation.tex
\begin{table*}[]
\centering
\setlength{\tabcolsep}{1pt}
\renewcommand{\arraystretch}{0.8} 
\footnotesize
\caption{Correlation coefficient values ($r_{pb}$) for LLMs' reasoning performance and complexity metrics (Table~\ref{metrics-complexity}).}
\vspace{-8pt}
\label{tab:correlation-analysis}
\resizebox{0.9\textwidth}{!}{
\begin{tabular}{|l|l|r|r|r|r|r|r|r|r|r|}
\hline
\multicolumn{1}{|c|}{\textbf{Task}}  &
  \multicolumn{1}{c|}{\textbf{Subject LLMs}} &
  \multicolumn{1}{c|}{\bm{$M_1$}} &
  \multicolumn{1}{c|}{\bm{$M_2$}} &
  \multicolumn{1}{c|}{\bm{$M_3$}} &
  \multicolumn{1}{c|}{\bm{$M_4$}} &
  \multicolumn{1}{c|}{\bm{$M_5$}} &
  \multicolumn{1}{c|}{\bm{$M_6$}} &
  \multicolumn{1}{c|}{\bm{$M_7$}} &
  \multicolumn{1}{c|}{\bm{$M_8$}} &
  \multicolumn{1}{c|}{\bm{$M_9$}} \\ \hline
 &
  Claude-Haiku-4.5 &
        -0.54 &
        -0.35 &
        -0.40 &
        -0.35&
        -0.44&
        -0.40&
        -0.30&
        -0.55&
        -0.32
  \\ \cline{2-11} 
 &
  Claude-Haiku-4.5(RD) &
        -0.59&
        -0.33&
        -0.38&
        -0.34&
        -0.51&
        -0.39&
        -0.28&
        -0.46&
        -0.30 
  \\ \cline{2-11} 
 &
  DeepSeek-V3.2 &
        -0.50&
        -0.32&
        -0.39&
        -0.33&
        -0.39&
        -0.38&
        -0.28&
        -0.52&
        -0.28
  \\ \cline{2-11} 
 &
  DeepSeek-V3.2(RD) &
       -0.53&
        -0.67&
        -0.42&
        -0.38&
        -0.45&
        -0.38&
        -0.29&
        -0.44&
        -0.34
  \\ \cline{2-11} 
 &
  Gemini-3-Pro &
        -0.50&
        -0.33&
        -0.57&
        -0.32&
        -0.62&
        -0.32&
        -0.25&
        -0.42&
        -0.30
  \\ \cline{2-11}
 &
  Gemini-3-Pro(LR) &
        -0.63&
        -0.32&
        -0.57&
        -0.35&
        -0.52&
        -0.35&
        -0.32&
        -0.42&
        -0.32
  \\ \cline{2-11}
 &
  GPT-5-mini &
        -0.55&
        -0.61&
        -0.55&
        -0.30&
        -0.37&
        -0.39&
        -0.26&
        -0.37&
        -0.26
  \\ \cline{2-11}
 &
  GPT-5-mini (LR) &
        -0.63&
        -0.56&
        -0.42&
        -0.36&
        -0.43&
        -0.38&
        -0.33&
        -0.45&
        -0.33
  \\ \cline{2-11}

   &
  CWM &
        -0.35&
        -0.28&
        -0.32&
        -0.28&
        -0.31&
        -0.32&
        -0.26&
        -0.37&
        -0.27
  \\ \cline{2-11}
\multirow{-6}{*}{\textbf{Input Prediction}} &
  CWM-Pretrain &
  \multicolumn{1}{l|}{-0.19} &
  \multicolumn{1}{l|}{-0.19} &
  \multicolumn{1}{l|}{-0.23} &
  \multicolumn{1}{l|}{-0.18} &
  \multicolumn{1}{l|}{-0.22} &
  \multicolumn{1}{l|}{-0.18} &
  \multicolumn{1}{l|}{-0.15} &
  \multicolumn{1}{l|}{-0.25} &
  \multicolumn{1}{l|}{-0.17} \\ \hline
 &
  Claude-Haiku-4.5 &
        -0.59&
        -0.33&
        -0.37&
        -0.32&
        -0.52&
        -0.35&
        -0.23&
        -0.51&
        -0.34 
  \\ \cline{2-11}
 &
  Claude-Haiku-4.5(RD) &
        -0.39&
        -0.56&
        -0.40&
        -0.33&
        -0.43&
        -0.33&
        -0.30&
        -0.44&
        -0.36
  \\ \cline{2-11} 
 &
  DeepSeek-V3.2 &
        -0.61&
        -0.33&
        -0.58&
        -0.33&
        -0.50&
        -0.31&
        -0.27&
        -0.41&
        -0.35
  \\ \cline{2-11} 
 &
  DeepSeek-V3.2(RD) &
       -0.36&
        -0.33&
        -0.39&
        -0.34&
        -0.42&
        -0.33&
        -0.26&
        -0.53&
        -0.35
  \\ \cline{2-11} 
 &
  Gemini-3-Pro &
        -0.54&
        -0.31&
        -0.57&
        -0.34&
        -0.42&
        -0.33&
        -0.26&
        -0.59&
        -0.36 
  \\ \cline{2-11} 
 &
  Gemini-3-Pro(LR) &
        -0.46&
        -0.34&
        -0.58&
        -0.34&
        -0.51&
        -0.37&
        -0.29&
        -0.41&
        -0.37
  \\ \cline{2-11} 

 &
  GPT-5-mini &
        -0.58&
        -0.62&
        -0.37&
        -0.31&
        -0.47&
        -0.32&
        -0.27&
        -0.40&
        -0.33 
  \\ \cline{2-11}
 &
  GPT-5-mini(LR) &
        -0.49&
        -0.31&
        -0.58&
        -0.35&
        -0.53&
        -0.34&
        -0.27&
        -0.42&
        -0.34
  \\ \cline{2-11} 
 &
  CWM &
        -0.24&
        -0.25&
        -0.26&
        -0.25&
        -0.27&
        -0.25&
        -0.21&
        -0.28&
        -0.26
  \\ \cline{2-11} 
\multirow{-6}{*}{\textbf{Output Prediction}} &
  CWM-Pretrain &
  \multicolumn{1}{l|}{-0.47} &
  \multicolumn{1}{l|}{-0.28} &
  \multicolumn{1}{l|}{-0.40} &
  \multicolumn{1}{l|}{-0.43} &
  \multicolumn{1}{l|}{-0.50} &
  \multicolumn{1}{l|}{-0.25} &
  \multicolumn{1}{l|}{-0.12} &
  \multicolumn{1}{l|}{
  -0.30} &
  \multicolumn{1}{l|}{-0.25} \\ \hline
\end{tabular}
}
\vspace{-5pt}
\end{table*}

%% file: Sections/3-6-RQ5.tex
\vspace{-5pt}
\subsection{RQ4: Ablation Study}
\label{rq:ablation}

\input{Tables/ablation}

\subsubsection{Prompting Strategies}

\input{Tables/ablation_categorization}
We evaluate two aspects of our prompting strategy.
For \emph{Structural Hints}, we removed the 
inputs/outputs structures from the prompt for both the ICL example and the query problem (e.g., \texttt{\small{\$ICL\_INPUT/ OUTPUT\_STRUCTUR\$}} and \texttt{\small{\$INPUT/OUTPUT\_STRUCTURE\$}} in Figure~\ref{fig:prompt-template} (a)). In the example of Figure~\ref{fig:complex-object} (c), instead of providing the JSON structure and variable names whose values should be predicted, we only ask LLM to predict the output of \texttt{\small{widget}}. For the \emph{CoT}, we removed both \emph{implicit} CoT (i.e., the sentence ``Execute the program step by step'' under the \texttt{\small{\#\#\#Instruction}}) and \emph{explicit} CoT (\texttt{\small{\$ICL\_CoT\$}}, where we teach the model to execute the code step by step) from the prompts. The results of the ablation study (Table~\ref{tab:ablation}) confirm the effectiveness of our prompt crafting strategy in helping better code reasoning. On average, \textbf{\emph{Structural Hints} improves $RS$ and $RS_{partial}$ by $5.36\%$ and $8.85\%$, respectively. The gain for LLMs is more notable on the \emph{HC} problems, where LLMs need to reason about a larger number of variables with complex types. Meanwhile, \emph{CoT} improves the performance of LLMs on  $RS$ and $RS_{partial}$ by $1.14\%$ and $1.69\%$, respectively.}

\subsubsection{Complexity Categorization Parameters}
There is a trade-off between the dataset size and the separation of complexity levels (Algorithm~\ref{alg:cap}). As shown in Table~\ref{tab:ablation-difficulty-algorithm}, increasing the cut-off threshold from $10\%$ (first row series) to $40\%$ (last row series) results in the selection of more problems for each category, but categories become less clearly separable. For each cut-off value, increasing the majority number results in a smaller but more clearly separate categorization. Algorithm~\ref{alg:cap} accounts for adjusting the values automatically depending on the collected problems, without any bias.



%% file: Tables/ablation.tex
\begin{table*}[]
\centering
\setlength{\tabcolsep}{1pt}
\footnotesize
\caption{Comparing the impact of \emph{Structural Hints} and \emph{CoT} on reasoning performance $\langle RS,RS_{partial} \rangle$.}
\vspace{-10pt}
\label{tab:ablation}
\resizebox{0.95\textwidth}{!}{
\begin{tabular}{|l|l|c|c|c|c|c|c|}
\hline
\multicolumn{1}{|c|}{\multirow{2}{*}{\textbf{Model}}} &
\multicolumn{1}{c|}{\multirow{2}{*}{\textbf{Ablation}}} &
  \multicolumn{3}{c|}{\textbf{Input Prediction\%}} &
  \multicolumn{3}{c|}{\textbf{Output Prediction\%}} \\ \cline{3-8} 

  \multicolumn{1}{|c|}{} &
\multicolumn{1}{c|}{} &
  \multicolumn{1}{c|}{\textbf{HC}} &
  \multicolumn{1}{c|}{\textbf{LC}} &
  \multicolumn{1}{c|}{\textbf{Total}} &
  \multicolumn{1}{c|}{\textbf{HC}} &
  \multicolumn{1}{c|}{\textbf{LC}} &
  \textbf{Total} \\ \hline
\multirow{2}{*}{\textbf{\haiku}} & -Structural Hints 
& $\langle 22.80, 45.08\rangle$
& $\langle 68.40, 79.43\rangle$ 
& $\langle 45.60, 62.26\rangle$
& $\langle 23.20, 24.40\rangle$  
& $\langle 72.40, 73.43\rangle$ 
& $\langle 47.80, 48.91\rangle$\\
& -CoT                   
& $\langle 28.80, 62.18\rangle$
& $\langle 69.20, 80.60\rangle$ 
& $\langle 49.00, 71.39\rangle$
& $\langle 36.00, 38.89\rangle$  
& $\langle 72.40, 73.44\rangle$
& $\langle 54.20, 56.67\rangle$ \\ \hline
\multirow{2}{*}{\textbf{\dsv}} & -Structural Hints 
& $\langle 20.40, 40.16\rangle$  
& $\langle 70.00, 80.56\rangle$ 
& $\langle 45.20, 60.36\rangle$ 
& $\langle 26.00, 28.23\rangle$ 
& $\langle 75.20, 75.96\rangle$ 
& $\langle 50.60, 52.09\rangle$ \\
& -CoT                   
& $\langle 30.40, 63.01\rangle$ 
& $\langle 68.40, 81.42\rangle$ 
& $\langle 49.40, 72.22\rangle$ 
& $\langle 29.05, 33.28\rangle$ 
& $\langle 54.40, 55.58\rangle$ 
& $\langle 41.73, 44.43\rangle$ \\ \hline
\multirow{2}{*}{\textbf{\geminip}} & -Structural Hints 
& $\langle 26.40, 45.81\rangle$ 
& $\langle 72.40, 84.23\rangle$ 
& $\langle 49.40, 65.02\rangle$ 
& $\langle 34.00, 40.06 \rangle$  
& $\langle 74.80, 76.36\rangle$ 
& $\langle 54.40, 58.21\rangle$ \\
& -CoT                   
& $\langle 33.60, 65.05\rangle$  
& $\langle 72.00, 83.34\rangle$ 
& $\langle 52.80, 74.19\rangle$ 
& $\langle 43.60, 50.07\rangle$ 
& $\langle 77.60, 79.52\rangle$ 
& $\langle 60.60, 64.80\rangle$ \\ \hline
\multirow{2}{*}{\textbf{\gfmini}} & -Structural Hints 
& $\langle 23.20, 40.82\rangle$  
& $\langle 68.00, 80.99\rangle$ 
& $\langle 45.60, 60.90\rangle$ 
& $\langle 24.80, 28.92\rangle$  
& $\langle 73.20, 75.25\rangle$ 
& $\langle 49.00, 52.09\rangle$ \\
& -CoT                   
& $\langle 30.40, 63.47\rangle$ 
& $\langle 67.60, 81.17\rangle$ 
& $\langle 49.00, 73.32\rangle$ 
& $\langle 38.00, 42.42\rangle$ 
& $\langle 76,80, 77.87\rangle$ 
& $\langle 57.40, 60.15\rangle$ \\ \hline
\multirow{2}{*}{\textbf{CWM}} & -Structural Hints 
& $\langle  12.94, 24.35\rangle$ 
& $\langle 40.40, 65.29\rangle$ 
& $\langle 26.26, 44.82\rangle$
& $\langle 10.00, 10.93 \rangle$ 
& $\langle 65.20, 65.73 \rangle$ 
& $\langle 37.60, 38.33\rangle$ \\
& -CoT                   
& $\langle  20.40, 38.67\rangle$ 
& $\langle 55.60, 69.09\rangle$ 
& $\langle 38.00, 53.88\rangle$ 
& $\langle 30.00, 33,73\rangle$ 
& $\langle 54.00, 55.49\rangle$ 
& $\langle 42.00, 44.61\rangle$\\ \hline

\multirow{2}{*}{\textbf{Average}} & -Structural Hints 
&$\langle 21.15, 39.24\rangle$ 
& $\langle 63.84, 78.10\rangle$ 
& $\langle 42.49, 58.67\rangle$ 
& $\langle 23.60, 26.51 \rangle$ 
& $\langle 72.16, 73.35 \rangle$ 
& $\langle 47.88, 49.39\rangle$ \\
& -CoT                   
& $\langle 28.72, 58.48 \rangle$ 
& $\langle 66.56, 79.12\rangle$ 
& $\langle 47.64, 68.80\rangle$ 
& $\langle 35.33, 39.88\rangle$ 
& $\langle 67.04, 68.38\rangle$ 
& $\langle 51.19, 54.13\rangle$
\\ \hline

\end{tabular}
}
\vspace{-16pt}
\end{table*}

%% file: Tables/ablation_categorization.tex
\begin{wraptable}{r}{0.37\textwidth}
\vspace{-30pt}
\caption{Impact of $T$ and $mojority$ parameters on categorization. Meaningless separations are highlighted with \textcolor[HTML]{FFCCC9}{\rule{0.9em}{0.9em}}.}
\vspace{-10pt}
\label{tab:ablation-difficulty-algorithm}
\centering
\scriptsize
\setlength{\tabcolsep}{5pt}        
\renewcommand{\arraystretch}{0.75}  
\begin{tabular}{|c|c|c|c|c|}
\hline
\textbf{Threshold} & \textbf{\#majority} & \textbf{\#HC} & \textbf{\#LC} \\
\hline
\multirow{5}{*}{\textbf{[0.1, 0.9]}}
  & 5 & 315 & 315 \\
  & 6 & 288 & 288 \\
  & 7 & 274 & 274 \\
  & 8 & 229 & 229 \\
  & 9 & 142 & 142 \\
\hline
\multirow{5}{*}{\textbf{[0.2, 0.8]}}
  & \cellcolor[HTML]{FFCCC9}5   & \cellcolor[HTML]{FFCCC9}684 & \cellcolor[HTML]{FFCCC9}684 \\
  & 6 & 544 & 544 \\
  & 7 & 487 & 487 \\
  & 8 & 405 & 405 \\
  & 9 & 208 & 208 \\
\hline
\multirow{5}{*}{\textbf{[0.25, 0.75]}}
  & \cellcolor[HTML]{FFCCC9}5   & \cellcolor[HTML]{FFCCC9}741 & \cellcolor[HTML]{FFCCC9}741 \\
  & \textbf{6} & \textbf{600} & \textbf{600} \\
  & 7 & 526 & 526 \\
  & 8 & 420 & 420 \\
  & 9 & 217 & 217 \\
\hline
\multirow{5}{*}{\textbf{[0.3, 0.7]}}
  & \cellcolor[HTML]{FFCCC9}5   & \cellcolor[HTML]{FFCCC9}817 & \cellcolor[HTML]{FFCCC9}817 \\
  & \cellcolor[HTML]{FFCCC9}6   & \cellcolor[HTML]{FFCCC9}662 & \cellcolor[HTML]{FFCCC9}662 \\
  & 7 & 561 & 561 \\
  & 8 & 479 & 479 \\
  & 9 & 254 & 254 \\
\hline
\multirow{5}{*}{\textbf{[0.4, 0.6]}}
  & \cellcolor[HTML]{FFCCC9}5   & \cellcolor[HTML]{FFCCC9}926 & \cellcolor[HTML]{FFCCC9}926 \\
  & \cellcolor[HTML]{FFCCC9}6   & \cellcolor[HTML]{FFCCC9}750 & \cellcolor[HTML]{FFCCC9}750 \\
  & \cellcolor[HTML]{FFCCC9}7   & \cellcolor[HTML]{FFCCC9}638 & \cellcolor[HTML]{FFCCC9}638 \\
  & 8 & 549 & 549 \\
  & 9 & 328 & 328 \\
\hline
\end{tabular}
\vspace{-8pt}
\end{wraptable}

%% file: Sections/3-5-RQ4.tex
\vspace{-5pt}
\subsection{RQ5: Other Reasoning Tasks}
\label{rq:additional}

\input{Tables/additional_results}

We also evaluate the extent to which the findings of input/output prediction tasks generalize to loop and branch prediction. Since not every reasoning problem in \name-lite contains a loop or conditional statement, especially within \emph{LC} problems, we performed this experiment on an equally sampled \emph{LC} and \emph{HC} problems ($100$ from each complexity level of \name-lite). 
Table~\ref{tab:additional_results} shows \textbf{a consistent trend} in the reasoning performance of LLMs about loops and branches: as we move from \emph{LC} to \emph{HC} problems, both $RS$ and $RS_{partial}$ values \textbf{drop substantially}. For loop prediction, $RS$ and $RS_{partial}$ decrease from $41.10\%$ to $20.20\%$ and $41.88\%$ to $24.64\%$, respectively. For branch prediction, $RS$ and $RS_{partial}$ decrease from $71.50\%$ to $22.90\%$ and $RS_{partial}$ and from $75.73\%$ to $48.75\%$, respectively. \textbf{$RS_{partial}$ values are consistently higher than $RS$}, by $2.61\%$ and $15.04\%$ for loop and branch prediction, respectively.

Loop prediction is harder to reason about in both complexity groups, likely because it requires the LLM to track a sequence of intermediate variable states; an error at any step can invalidate the final prediction. For branch prediction, LLMs achieve performance comparable to input/output prediction on \emph{LC} problems, as these programs contain only a few simple branches. \textbf{Given that the decisions are binary and usually require predicting a single predicate on \emph{LC}, a random guess may also be in favor of LLMs~\cite{liu2025assessing,chen2024reasoning}}. However, when moving to \emph{HC} problems, branch prediction becomes substantially more difficult: \emph{HC} programs contain more conditional statements, particularly those nested within other conditional statements or loops, introducing non-trivial reasoning challenges and preventing inflated performance due to lucky guesses. 

Consistent with input/output prediction (\S \ref{subsub:call-chain}), call chain size shows a moderate to strong negative correlation with LLM success on loop and branch prediction ($r_{pb}=-0.61$ and $-0.47$, respectively). Similarly, enabling or increasing the reasoning effort consistently improves the reasoning performance. In loop prediction, the average gains are $5.90\%$ in $RS$ and $5.76\%$ in $RS_{partial}$ over non-reasoning/low-reasoning counterparts. In branch prediction, the corresponding improvements are $7.80\%$ and $7.46\%$, respectively. Similar to input/output prediction (\S \ref{subsub:training-strategies}), non-/low-reasoning models, albeit to a negligible extent, can solve unique problems in loop/branch prediction compared to their reasoning-enabled counterparts. Finally, we augment the reasoning failure categorization (\S \ref{rq:reasoning-failure}) with the loop prediction mistakes (under \emph{Loop Variable Misunderstanding} and \emph{Loop Iterable Misunderstanding}) and branch prediction mistakes (under \emph{Incorrect Branch Prediction}).

%% file: Tables/additional_results.tex
\begin{table*}[]
\setlength{\tabcolsep}{1pt}
\footnotesize
\caption{Performance of LLMs $\langle RS,RS_{partial} \rangle$ on loop and branch prediction using \name-lite.}
\vspace{-8pt}
\label{tab:additional_results}
\resizebox{0.95\textwidth}{!}{
\begin{tabular}{|l|lll|lll|}
\hline
\multirow{2}{*}{\textbf{Subject LLMs}} &
  \multicolumn{3}{c|}{\textbf{Loop Prediction\%}} &
  \multicolumn{3}{c|}{\textbf{Branch Prediction\%}} \\ \cline{2-7} 
 &
  \multicolumn{1}{c|}{\textbf{HC}} &
  \multicolumn{1}{c|}{\textbf{LC}} &
  \multicolumn{1}{c|}{\textbf{Total}} &
  \multicolumn{1}{c|}{\textbf{HC}} &
  \multicolumn{1}{c|}{\textbf{LC}} &
  \multicolumn{1}{c|}{\textbf{Total}} \\ \hline
\textbf{Claude-Haiku-4.5} &
  \multicolumn{1}{l|}{$\langle 23.00, 25.92\rangle$} &
  \multicolumn{1}{l|}{$\langle 41.00, 41.50 \rangle$} &
  $\langle 32.00, 33.71 \rangle$ &
  \multicolumn{1}{l|}{$\langle 26.00, 55.41 \rangle$} &
  \multicolumn{1}{l|}{$\langle 79.00, 83.62\rangle$} &
  $\langle 52.50, 69.50 \rangle$ \\ \hline
\textbf{DeepSeek-V3.2} &
  \multicolumn{1}{l|}{$\langle 24.00, 31.81\rangle$} &
  \multicolumn{1}{l|}{$\langle 50.00, 51.00\rangle$} &
  $\langle 37.00, 41.41 \rangle$ &
  \multicolumn{1}{l|}{$\langle 26.00, 55.00 \rangle$} &
  \multicolumn{1}{l|}{$\langle 76.00, 79.82\rangle$} &
  $\langle 51.00, 67.14  \rangle$ \\ \hline
\textbf{Gemini-3-Pro} &
  \multicolumn{1}{l|}{$\langle 26.00, 30.25 \rangle$} &
  \multicolumn{1}{l|}{$\langle 54.00, 54.30\rangle$} &
  $\langle 40.00, 42.28 \rangle$ &
  \multicolumn{1}{l|}{$\langle 30.00, 52.00 \rangle$} &
  \multicolumn{1}{l|}{$\langle 80.00, 83.95 \rangle$} &
  $\langle 55.00, 68.98\rangle$ \\ \hline
\textbf{GPT-5-mini} &
  \multicolumn{1}{l|}{$\langle 23.00, 27.06 \rangle$} &
  \multicolumn{1}{l|}{$\langle 48.00, 48.50\rangle$} &
  $\langle 35.50, 37.78\rangle$ &
  \multicolumn{1}{l|}{$\langle 27.00, 53.00\rangle$} &
  \multicolumn{1}{l|}{$\langle 77.00, 82.21\rangle$} &
  $\langle 52.00, 67.61\rangle$ \\ \hline
\textbf{CWM} &
  \multicolumn{1}{l|}{$\langle 17.00, 19.52\rangle$} &
  \multicolumn{1}{l|}{$\langle 30.00, 31.50\rangle$} &
  $\langle 23.50, 25.51  \rangle$ &
  \multicolumn{1}{l|}{$\langle 22.00, 42.62 \rangle$} &
  \multicolumn{1}{l|}{$\langle 68.00, 72.62\rangle$} &
  $\langle 45.00, 57.62 \rangle$ \\ \hline
\textbf{Claude-Haiku-4.5 (RD)} &
  \multicolumn{1}{l|}{$\langle 21.00, 23.78\rangle$} &
  \multicolumn{1}{l|}{$\langle 33.00, 34.00\rangle$} &
  $\langle 27.00, 28.89\rangle$ &
  \multicolumn{1}{l|}{$\langle 26.00, 55.00 \rangle$} &
  \multicolumn{1}{l|}{$\langle 79.00, 82.70\rangle$} &
  $\langle 52.50, 68.85\rangle$ \\ \hline 
  
  \textbf{DeepSeek-V3.2 (RD)} &
  \multicolumn{1}{l|}{$\langle 21.00, 28.25\rangle$} &
  \multicolumn{1}{l|}{$\langle 43.00, 43.50\rangle$} &
  $\langle 32.00, 35.88\rangle$ &
  \multicolumn{1}{l|}{$\langle 23.00, 50.32 \rangle$} &
  \multicolumn{1}{l|}{$\langle 66.00, 68.53\rangle$} &
  $\langle 44.50, 59.43\rangle$ \\ \hline

\textbf{Gemini-3-Pro (LR)} &
  \multicolumn{1}{l|}{$\langle 20.00, 27.33 \rangle$} &
  \multicolumn{1}{l|}{$\langle 43.00, 44.50\rangle$} &
  $\langle 31.50, 35.92\rangle$ &
  \multicolumn{1}{l|}{$\langle 17.00, 49.72\rangle$} &
  \multicolumn{1}{l|}{$\langle 77.00, 82.62\rangle$} &
  $\langle 47.00, 66.17 \rangle$ \\ \hline

\textbf{GPT-5-mini (LR)} &
  \multicolumn{1}{l|}{$\langle 18.00, 21.64\rangle$} &
  \multicolumn{1}{l|}{$\langle 44.00, 44.00\rangle$} &
  $\langle 31.00, 32.82 \rangle$ &
  \multicolumn{1}{l|}{$\langle 22.00, 51.00 \rangle$} &
  \multicolumn{1}{l|}{$\langle 76.00, 80.60\rangle$} &
  $\langle 49.00, 65.80\rangle$ \\ \hline 
  \textbf{CWM (Pretrain)} &
  \multicolumn{1}{l|}{$\langle 9.00, 10.83 \rangle$} &
  \multicolumn{1}{l|}{$\langle 25.00, 26.00\rangle$} &
  $\langle 17.00, 18.42 \rangle$ &
  \multicolumn{1}{l|}{$\langle 10.00, 23.40 \rangle$} &
  \multicolumn{1}{l|}{$\langle 37.00, 41.17 \rangle$} &
  $\langle 23.50, 32.29\rangle$ \\ \hline \hline

  \textbf{Average} &
  \multicolumn{1}{l|}{$\langle 20.20, 24.64 \rangle$} &
  \multicolumn{1}{l|}{$\langle 41.10, 41.88\rangle$} &
  $\langle 30.65, 33.26 \rangle$ &
  \multicolumn{1}{l|}{$\langle 22.90, 48.75 \rangle$} &
  \multicolumn{1}{l|}{$\langle 71.50, 75.73 \rangle$} &
  $\langle 47.20, 62.24\rangle$ \\
  
  \hline
\end{tabular}
}
\vspace{-15pt}
\end{table*}

%% file: Sections/Related-Work.tex
\vspace{-5pt}
\section{Related Work}

\name is related to studies on code reasoning tasks~\cite{gu2024cruxeval,liu2024codemind,chen2024reasoning,beger2025coconut,kopelexecution,le2025can,yang2025evaluating,sun2025l0, gautam2025refactorbench, bolet2025counting, pujar2025code, bolet2025counting}. 
\crux reasoning tasks are input and output prediction. \codemind proposes three tasks (independent execution reasoning, dependent execution reasoning, and specification reasoning) to evaluate LLMs' inductive code reasoning abilities. \reval~\cite{chen2024reasoning} studies runtime behavior in the reasoning process through code coverage prediction, program state prediction, execution path prediction, and output prediction. Mofia et al.~\cite{la2025code} show that executing code can act as a proxy for naturalistic tasks, including value exchange, repetitive computation, and object ranking. \coco~\cite{beger2025coconut} challenges LLMs to generate a trace of line numbers executed by the program for a given set of inputs. EquiBench~\cite{wei2025equibench} assesses LLMs’ understanding of code execution semantics by judging whether two programs produce equivalent outputs for every possible input. CES~\cite{liu2025assessing} integrates output prediction with reasoning about key decision points in the program into a single prompt and evaluates LLMs’ reasoning coherence and consistency. HoarePrompt~\cite{bouras2025hoareprompt} proposes an iterative prompting strategy that harnesses LLMs’ code reasoning to verify program correctness against the natural language specification. CoRe~\cite{xie2025core} assesses LLMs' code reasoning capability on static analysis tasks, including code dependency reasoning, control dependency reasoning, and information flow reasoning.
\textbf{None of the prior studies involve real-world reasoning problems, failing to challenge the reasoning abilities of LLMs properly as discussed in this paper.}

Two closely related studies are EXE~\cite{kopelexecution} and CodeSense~\cite{roy2025codesense} that evaluate LLM code reasoning on problems from open-source Python repositories. \textbf{EXE relies on LLMs to generate test inputs, limiting them to simple and primitive variable types}. CodeSence collects input-output pairs from actual execution trace logs, but \textbf{rules out those with non-primitive data types.}
To better preserve the complexity of real-world code inputs, \name collects actual runtime inputs through existing tests. 
It can serialize any complex custom type into a readable JSON-like format. \textbf{These studies also assume that a problem from a real-world project is always complex, which we show is not necessarily true, as some \name and \name-lite LC instances are from real-world projects} (Figure~\ref{fig:difficulty-level}). \name also follows a systematic approach to categorize the reasoning benchmarks into two complexity groups, not done by any prior studies.

%% file: Sections/Conclusion.tex
\vspace{-8pt}
\section{Threats to Validity}
\label{sec:threats}

To mitigate the threat of generalizing results to other experimental settings (other models, benchmarks, and tasks), our selected models cover both open-source and API access models, as well as general and reasoning models. 
Concerning new benchmarks, \name pipeline is generic, i.e., it can convert any Python project with executable test suites into reasoning problems. The rigorous design of complexity-level categorization of problems makes \name adaptive to new problems and new LLM capabilities, as discussed in the paper.
As a proof of concept, this paper shows that the code reasoning abilities of LLMs in four widely used tasks of input, output, loop, and branch prediction can be significantly challenged under real-world settings. \name currently focuses on Python, as the majority of the programming datasets widely use it. Focusing on one language also enables systematic control over complexity metrics and avoids language-induced variance. 

LLMs are inherently non-deterministic. One way to account for non-determinism is repeating the experiments $k$ times, and reporting the $RS@k$ results.
However, a similar computation is not possible for $RS_{partial}$ due to the inherent dependency between attributes of complex objects. Given that in practice, users of LLMs do not repeatedly prompt LLMs and pick the correct solution, if it exists, this design decision does not pose a threat to the validity of the results. 


\vspace{-8pt}
\section{Conclusion}
\label{sec:conclusion}

In this paper, we introduce \name, in which reasoning problems are categorized into two complexity groups. We evaluate ten reasoning and general LLMs on input, output, loop, and branch prediction tasks using \name-lite problems and observe a substantial performance drop from $LC$ to $HC$ programs, revealing that existing benchmarks may overstate LLMs' reasoning capabilities. For future work, we plan to support more programming languages and cross-analysis between code reasoning and other programming tasks.